\documentclass[a4paper,fleqn,usenatbib]{mnras}
\usepackage[T1]{fontenc}
\usepackage{ae,aecompl}
\usepackage{graphicx}   
\usepackage{amsmath}    
\usepackage{amssymb}    

\newcommand{\Msun}{$M_\odot$}
\newcommand{\Rsun}{$R_\odot$}

\title[Shock breakouts from red supergiants: analytical and numerical predictions]
{Shock breakouts from red supergiants: \\ analytical and numerical predictions}

\author[A. Kozyreva et al.]
{Alexandra~Kozyreva$^{1}$\thanks{E-mail: sasha@mpa-garching.mpg.de}, Ehud Nakar$^{2}$,
Roni Waldman$^{3}$, 
\newauthor
Sergei~Blinnikov$^{4,5,6}$, Petr Baklanov$^{4,5,7}$
\\
$^{1}$Max-Planck-Institut f\"ur Astrophysik, Garching bei M\"unchen, 85748, Germany\\
$^{2}$The Sackler School of Physics and Astronomy, Tel Aviv University, Tel Aviv, 6997212, Israel\\
$^{3}$Racah Institute of Physics, The Hebrew University, Jerusalem 91904, Israel\\
$^{4}$Space Research Institute (IKI), 84/32 Profsoyuznaya, Moscow, 117997, Russia\\
$^{5}$NRC ``Kurchatov Institute'' -- ITEP, Moscow, 117218, Russia \\
$^{6}$Kavli IPMU (WPI), University of Tokyo, Kashiwa, Chiba 277-8583, Japan\\
$^{7}$National Research Nuclear University Moscow Engineering Physics Institute, Moscow, 115409, Russia
}

\date{Accepted XXX. Received YYY; in original form ZZZ}
\pubyear{2019}

\begin{document}
\label{firstpage}
\pagerange{\pageref{firstpage}--\pageref{lastpage}}
\maketitle

\begin{abstract}

Shock breakout (SBO) signal is the first signature of the supernova explosion
apart from gravitational waves and neutrinos.
Observational properties of SBO, such as bolometric luminosity and colour
temperature, connect to the supernova progenitor and explosion parameters.
Detecting SBO or SBO-cooling will constrain the progenitor and explosion models of
collapsing stars. In the light of recently launched eROSITA telescope, the
rate for detection of SBO is a few events during a year.
In the current study, we examine the analytic formulae derived by
Shussman\,et\,al.\,(2016). We use four red supergiant models from their study, while running
explosions with the radiation hydrodynamics code \verb|STELLA|. We conclude
that there is a good agreement between analytic and numerical approaches for 
bolometric luminosity and colour temperature during SBO. 
The analytic formulae for the SBO signal based on the
global supernova parameters can be used instead of running
time-consuming numerical simulations.
We define spectral range where analytic formulae 
for the SBO spectra are valid. We
provide improved analytical expression for the SBO spectral energy distribution. 
We confirm dependence of colour temperature on radius derived by analytical
studies and suggest to use early time observations to confine the progenitor radius.
Additionally we show the prediction for the SBO
signal from red supergiants as seen by eROSITA instrument.

\end{abstract}


\section{Introduction} 
\label{sect:intro}

Nowadays, it is become possible to detect supernovae at very early phase and resolve
the light curve behavior very soon after the supernova explosion. With
short-cadence surveys observers are able to catch the earliest light
indicating the explosion \citet{2017ApJ...837L...2A}. The first
electro-magnetic signature of the explosion of a massive star is the shock breakout (SBO) event,
i.e. the emergence of the shock on the surface of the progenitor star.
However, the SBO peaks in extreme ultraviolet, while the majority of surveys specified in
longer wavelengths. So far, there are only few SBO events detected 
\citep{2008ApJ...683L.131G,2008Sci...321..223S,2008Natur.453..469S,2009ApJ...702..226M}.
The difficulty consists in the short duration of the SBO, spanning minutes
for compact progenitors (fraction of solar radius to hundred solar radii)
to hours for extended progenitors, e.g. red
supergiants (hundreds to thousands solar radii).


One of the most important aspects of the SBO is that 
the SBO properties like bolometric luminosity and temperature depend on the
progenitor radius and indirectly depend on ejecta mass and
explosion energy. Hence, detecting SBOs from normal core-collapse supernovae
(CCSNe) which originate from the explosion of red supergiants \citep{2009ARA&A..47...63S}
will help to constrain progenitor and explosion properties. Estimates for
the unknown characteristics like progenitor radius and explosion energy from
the SBO phase serve as an additional constrain to the first-order
approximations for supernova parameters from 
\citet{1985SvAL...11..145L,1993ApJ...414..712P} 
and latest corrected relationships 
\cite{2019ApJ...879....3G} and will break the parameter degeneracy
of the explosion models \citep{2019A&A...625A...9D}.
One of the possible ways to derive the required parameters is to compare the
SBO properties to the analytic formulae or numerical simulations. While
detailed numerical simulations lead to a more exact model, analytical
approach provides the simplest procedure to get approximate values for
the progenitor.

Amongst others, \citet{2010ApJ...725..904N}, \citet{2011ApJ...728...63R}, and later \citet{Shussman2016} derive
the analytic prescription for the early light curve of SBO, the color temperature, and
spectra. \citet{Shussman2016} calibrate the analytic relations with the numerical simulations based on
120 red supergiant models. 
While solving the reverse problem using the analytic formulae, one may
estimate progenitor and explosion parameters from the observed light curve, temperature, and
spectra without running time-consuming numerical simulations of stellar evolution and
explosion models. 

In the present study, we run a few models from the \citet{Shussman2016} set
using more sophisticated multi-group radiation hydrodynamics code (\verb|STELLA|) 
and confirm reasonable agreement to their analytic prescription.
We describe the models and the method in Section~\ref{sect:method}, compare
our numerical results with the analytic formulae in
Section~\ref{sect:results}, analyse the spectral range in which analytic
expressions provide reliable fits, and summarise the current study in
Section~\ref{sect:conclusions}, while offering the improved relations for
the SBO signature.

\section[Models and Method]{Models and Method} 
\label{sect:method}

\begin{figure}
\centering
\includegraphics[width=0.5\textwidth]{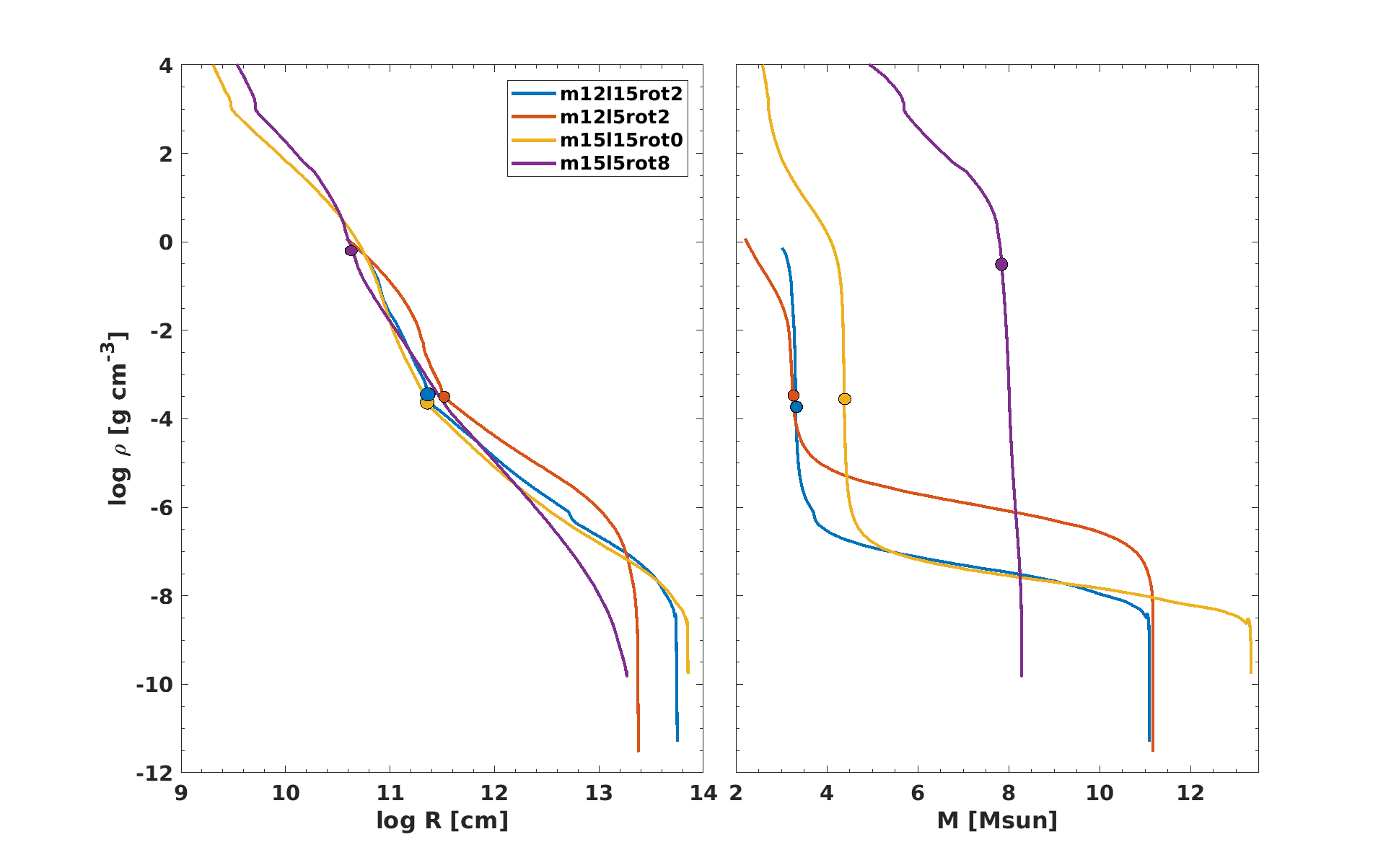}
\caption{Density structure of the red supergiant models in the study. Dots on each
curve indicate the bottom of hydrogen-rich envelope.}
\label{figure:all_RhoR}
\end{figure}

For our study we selected four core-collapse progenitor models from \citet{Shussman2016}. 
Among 120 progenitors from their study, we chose three typical progenitors that
cover almost the entire range  of radii (345-1024~\Msun{}) and one progenitor
that is extreme, with relatively smaller radius and atypical density
structure.
These are: m12l15rot2, m12l5rot2, m15l15rot0, m15l5rot8 -- all computed with
\verb|MESA|\footnote{Modules for Experiments in Stellar Astrophysics
\url{http://mesa.sourceforge.net/}
\citep{2011ApJS..192....3P,2013ApJS..208....4P,2015ApJS..220...15P,2018ApJS..234...34P}.}. The
Table~\ref{table:models}
contains some characteristics of our models such as radius, final and ejecta mass,
parameters used for stellar evolution calculations, such as
a parameter of Ledoux convection and 
rotation in units of Keplerian equatorial velocity.
All stellar models are at solar metallicity, calculated with 
semiconvection parameter 0.1 and
exponential overshoot formula with the parameter of 0.008. In
Figure~\ref{figure:all_RhoR}, we show
the density structure of all four models. The radius varies in a wide range
from 268~$R_\odot${} to 1024~$R_\odot${}, while density gradient stays the
same for all models except the model m15l5rot8, which has a sharper slope of
density in the hydrogen-rich envelope.

\begin{table}
\caption[RSG models]
{Characteristics of the red supergiants used in the present study.
Radius is in solar radii. Final and ejecta masses are in solar masses.  
``Convection'' stands for the parameter of Ledoux convection.
``Rotation'' means equatorial velocity in units of Keplerian velocity.}
\label{table:models}
\begin{center}
\begin{tabular}{l|r|c|l|l|c}
model  & Radius     & $M_{final}$ & $M_{ej}$  & Convection & Rotation \\
name   & [\Rsun{}]  & [\Msun{}]   & [\Msun{}] &            &           \\
\hline
m12l15rot2 & 812 & 11.09 & 8.09 &1.5& 0.2\\
m12l5rot2  & 345 & 11.17 & 9.15 & 5 & 0.2\\
m15l15rot0 & 1024& 13.33 & 11.33&1.5& --\\
m15l5rot8  & 268 & 8.3   & 6.1  &5  & 0.8\\
\end{tabular}
\end{center}
\end{table}

For the explosion and post-explosion evolution we used the radiation hydrodynamics
code \verb|STELLA| \citep{2006A&A...453..229B}. The explosion is initiated
in \verb|STELLA| as a thermal bomb. 1.1~foe of thermal energy 
(1~foe\,$\equiv 10^{\,51}$\,erg) is
injected at the bottom of the inner boundary in the inner 0.3~\Msun{} of
the supernova ejecta, on top of the region which is considered to collapse
into a proto-neutron star. By the end of simulations (1-2~days after the
explosion), the final kinetic energy is about 1~foe for all models.
In the model m12l15rot2 we truncated the inner 4~\Msun{} to
simplify calculations to avoid some computational issues. Nevertheless the
exact inner structure does not affect the resulting shock breakout
properties and does not influence the conclusions of the present study. In
general, the simulations are carried out with the standard opacity treatment, and with the
standard numerical parameter set used for the majority of \verb|STELLA|
studies. We pay special attention to the parameter BQ which is designed to resolve SBO
evolution according to \citet[][]{2011ApJS..193...20T}. The standard value is
BQ\,=\,1, while we set BQ\,=\,0 which is needed specifically for the SBO simulations.
The parameter is responsible for the
efficiency of energy conversion from kinetic energy of the shock into
thermal energy \citep[see detailed explanation in
Section~2.5.2.3][]{2013PhDTMoriya}. The value BQ\,=\,0
represents the idealised 100\%-efficiency of energy conversion and corresponds to Rankine-Hugoniot
conditions. On top of that, BQ ``smears'' velocity and density
gradients among a few meshes \citep[see Appendix in][]{1998ApJ...496..454B} 
to mimic multi-dimensional effects which develop during the ejecta expansion \citep{2016MNRAS.459.2188B}.
The ejecta is spherical and ``compact'' at the moment of SBO 
relative to the later supernova phases. Later different multi-dimensional instabilities
appear, and this makes
conversion of kinetic energy less efficient. Keep running simulations for
normal supernovae with
BQ=0 on the longer time scale provides less physical results \citep[see also
Fig.~4.13, Section~4.3.3.2 in ][]{2013PhDTMoriya}.

The spectral energy distribution (SED) is computed in the wavelength
spanning from 1\,\AA{} to 50,000\,\AA{}. The frequency range is divided on
100 bins logarithmically, in which the radiative transfer equations are
solved at every time step. The final bolometric light curve is integrated over
the spectra. Colour temperature is estimated as a black body temperature via
the least-square method.

We emphasize that in the present analysis we do multi-group radiation simulations
coupled with hydrodynamical evolution of the ejecta which is the
next step in the complexity after gray-opacity calculations done by \citet{Shussman2016}.
Of course, there are a number of published SBO studies, e.g. 
\citet[][ one-temperature simulations]{1992ApJ...393..742E},
\citet[][ STELLA]{2011ApJS..193...20T} and 
\citet[][ STELLA]{2013MNRAS.429.3181T},
\citet[][ gray-opacity, diffusion approximation, SNEC]{2016ApJ...829..109M},
\citet[][ gray-opacity, CASTRO]{2017ApJ...845..103L},
\citet[][ HERACLES/non-LTE CMFGEN]{2017A&A...605A..83D} and others.

\section[Comparison between numerical and analytical approach]{Comparison between numerical and analytical approach} 
\label{sect:results}

\subsection[Bolometric light curves and color temperature]{Bolometric light curves and color temperature}
\label{subsect:LbolTcol}

\begin{figure}
\centering
\includegraphics[width=0.5\textwidth]{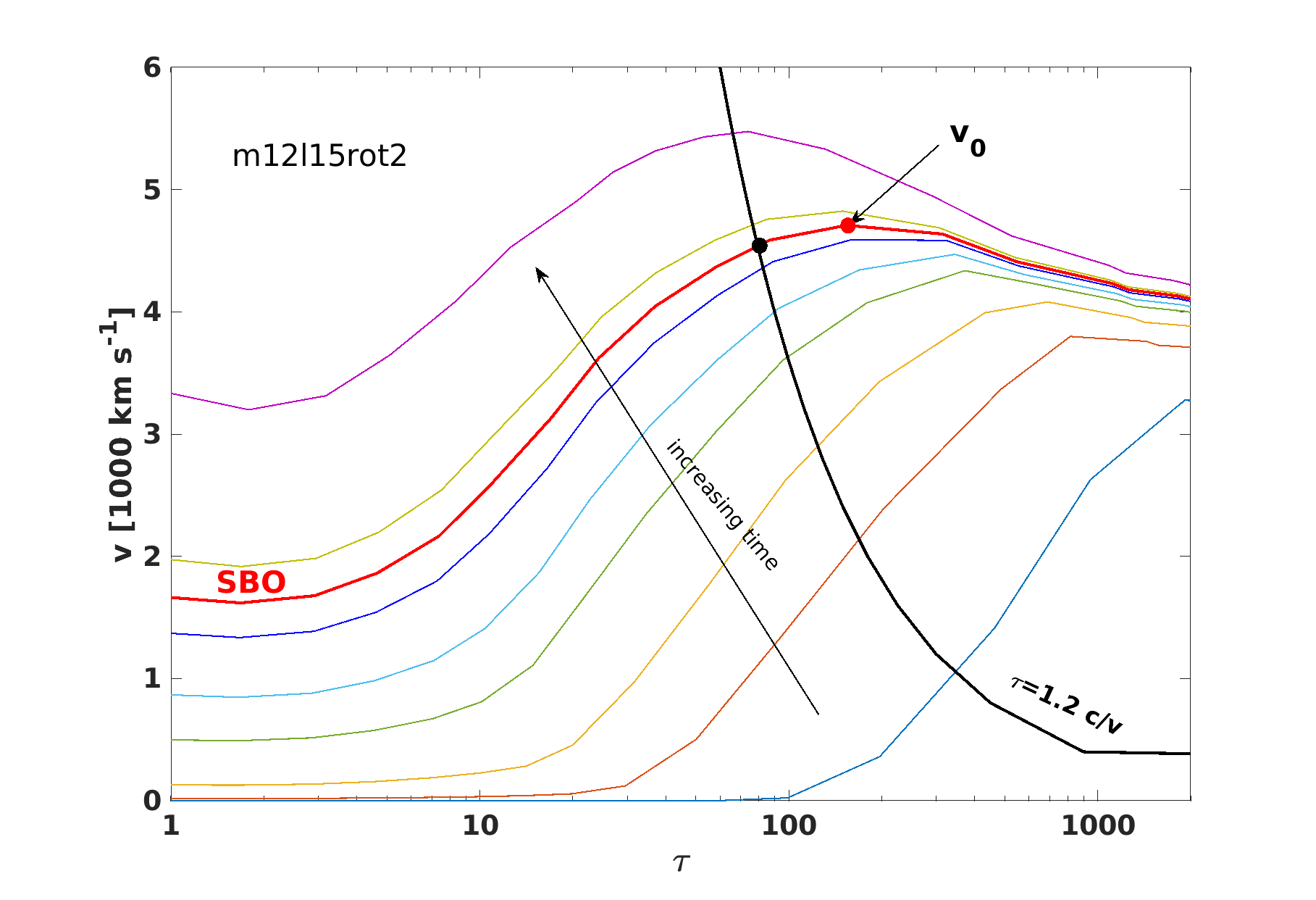}
\caption{The choice for $v_0,\,\rho_0$ of the breakout shell required for
calculating SBO properties.}
\label{figure:VeloTau}
\end{figure}

\begin{figure*}
\centering
\includegraphics[width=0.5\textwidth]{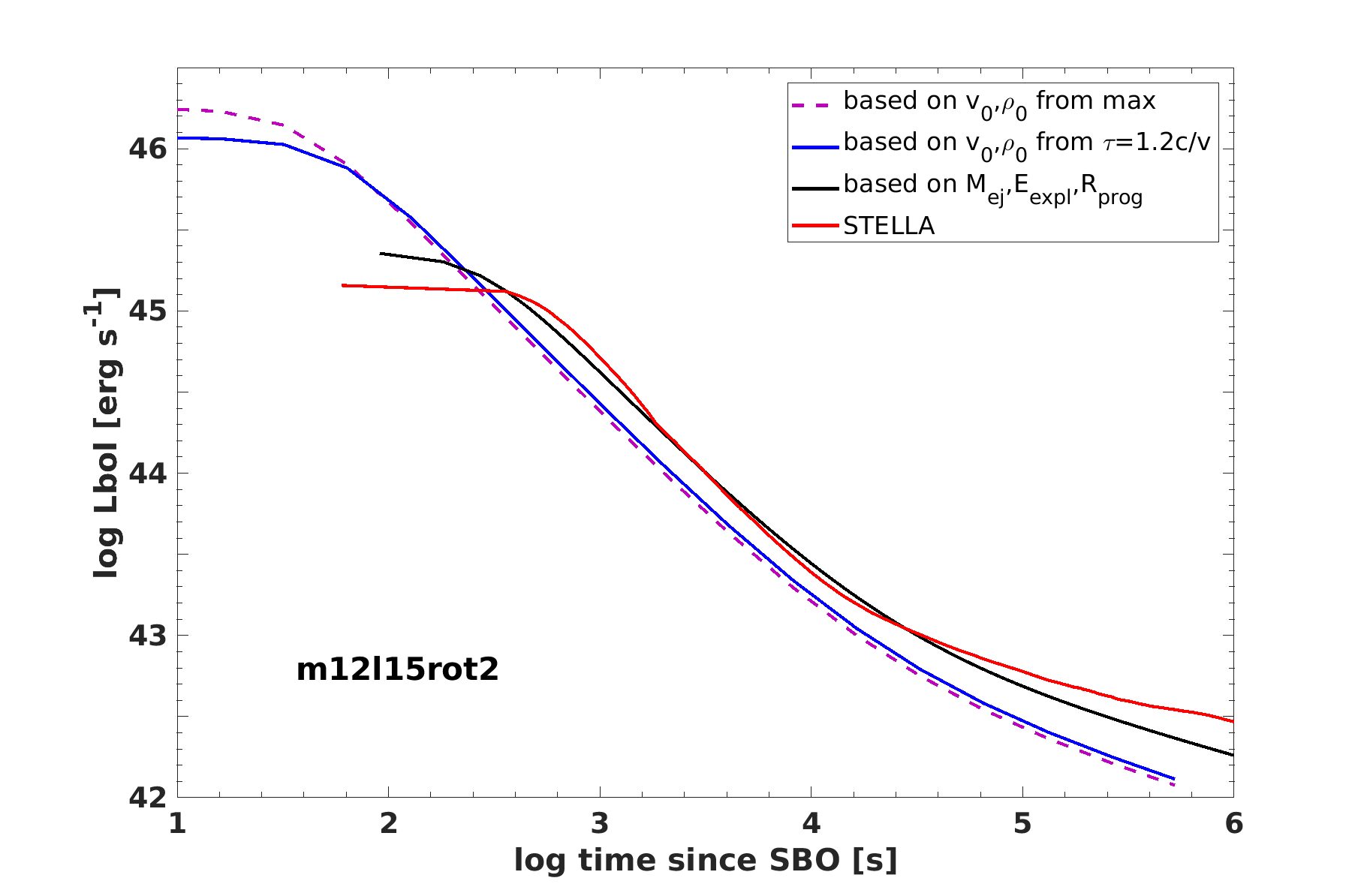}~
\includegraphics[width=0.5\textwidth]{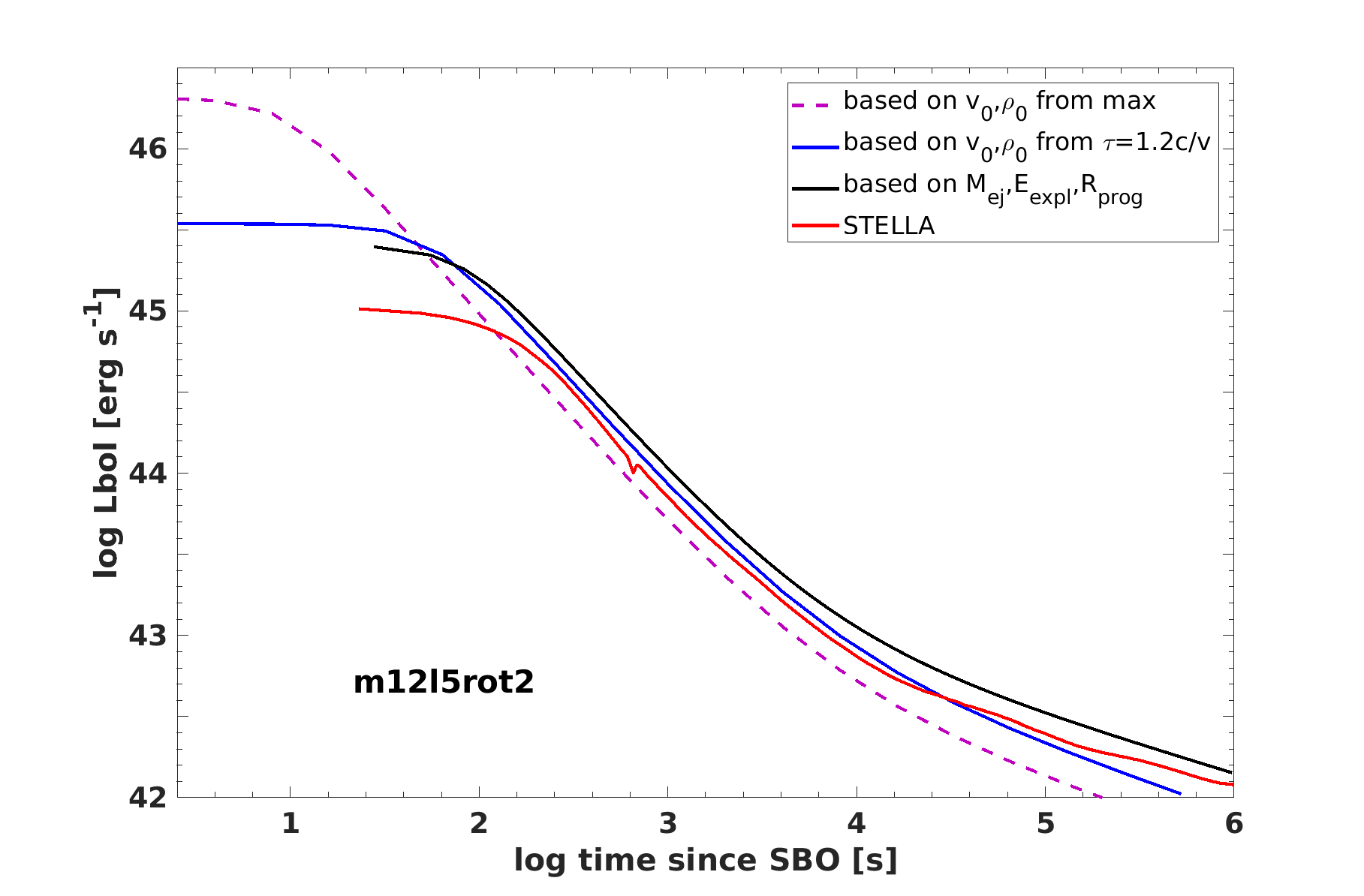}\\
\includegraphics[width=0.5\textwidth]{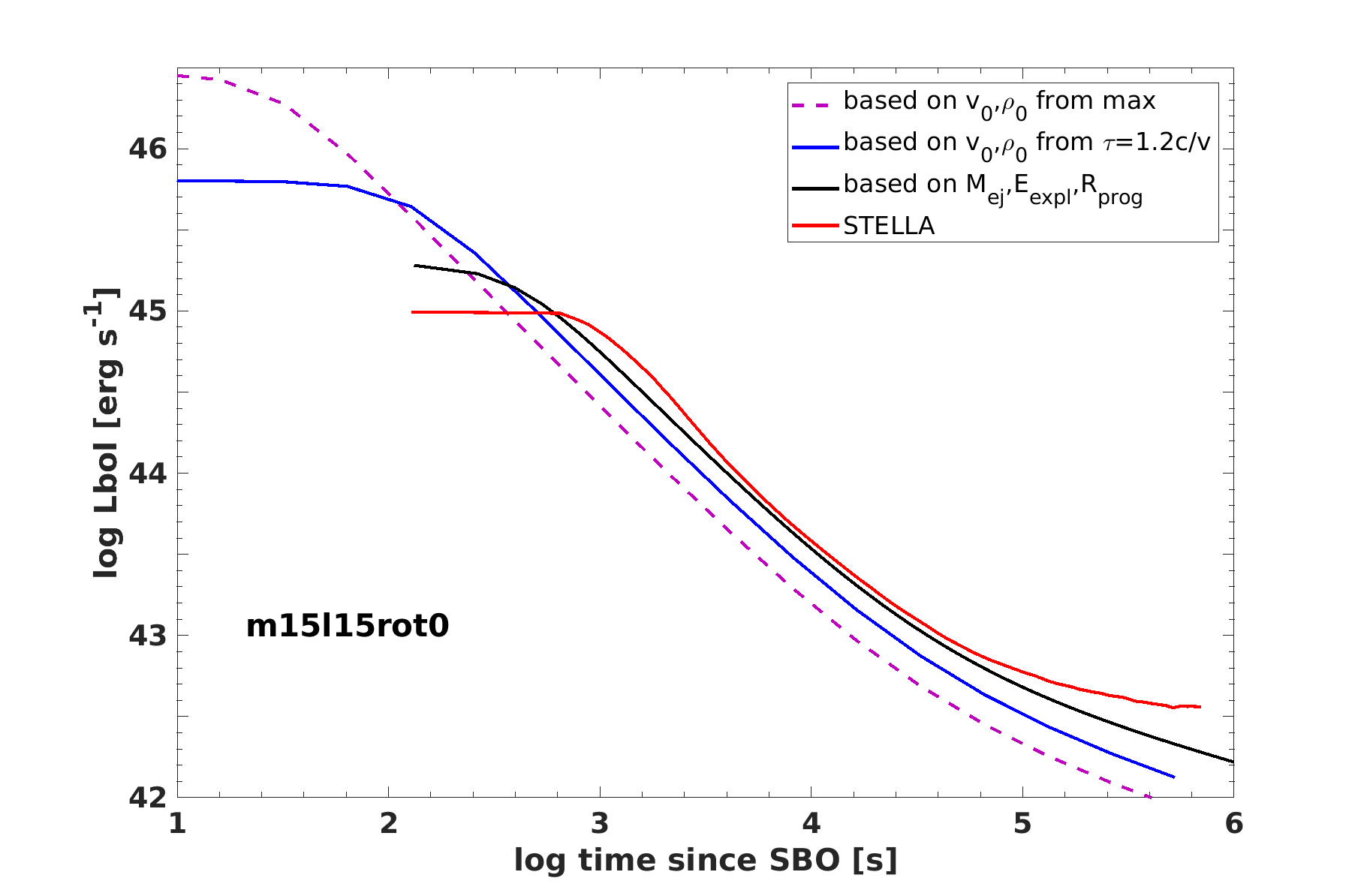}~
\includegraphics[width=0.5\textwidth]{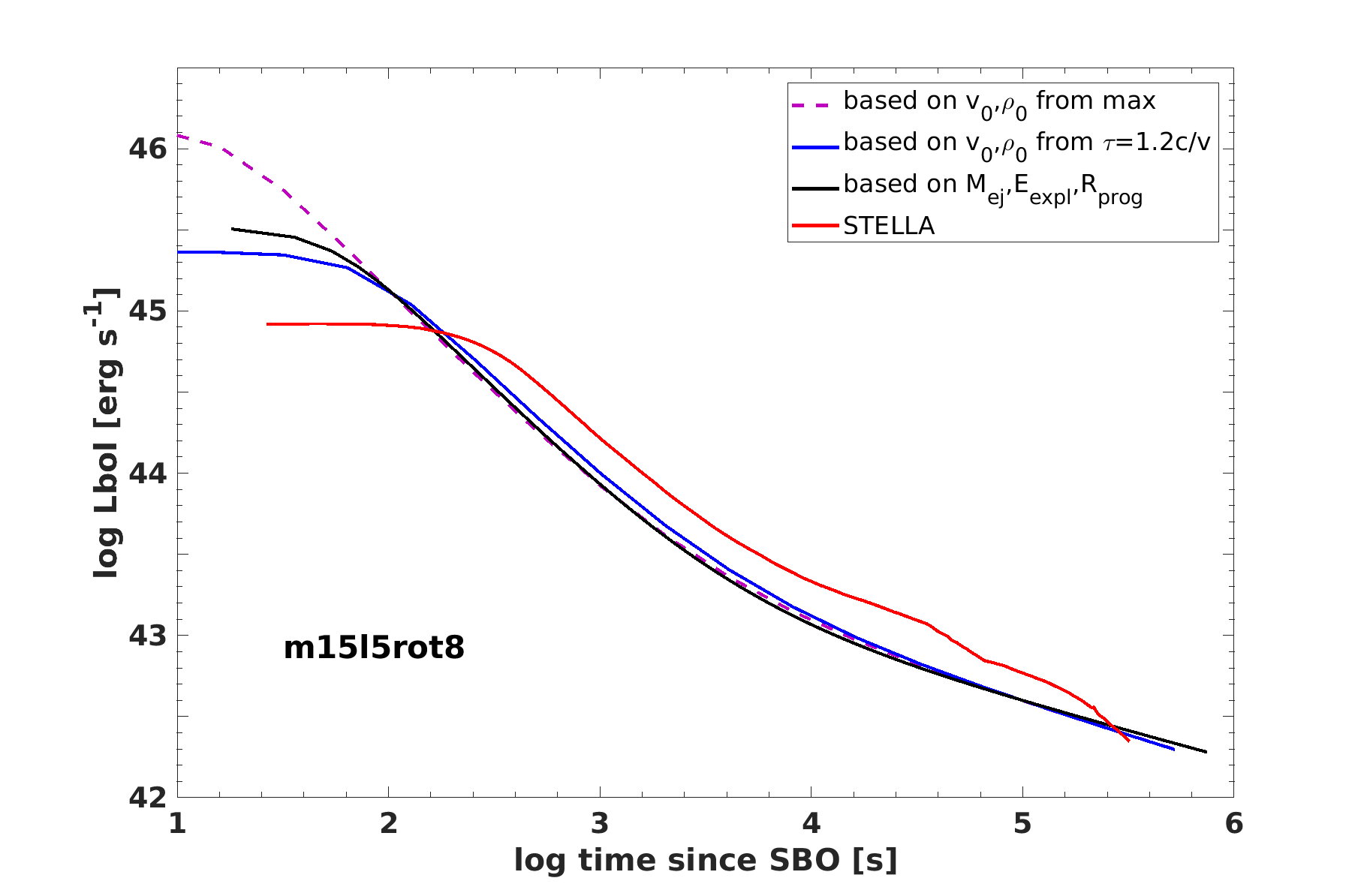}\\
\caption{SBO bolometric light curves for red supergiant models m12l15rot2, m12l5rot2,
m15l15rot0, and m15l5rot8. The dashed curve represents bolometric luminosity
from Eq.~33 (see Appendix~\ref{appendix:append}) based on breakout shell parameters $v_0, \rho_0$ at
maximum shock velocity. The blue curve is calculated with the same Eq.~33 based on $v_0, \rho_0$ at
the point where $\tau=1.2\,c/v_0$. The black curve is calculated with Eq.~39
based on global supernova properties $M_{ej}, E_{expl},
R_{prog}$. See discussion in the text.}
\label{figure:all_Lbol}
\end{figure*}

\begin{figure*}
\centering
\includegraphics[width=0.5\textwidth]{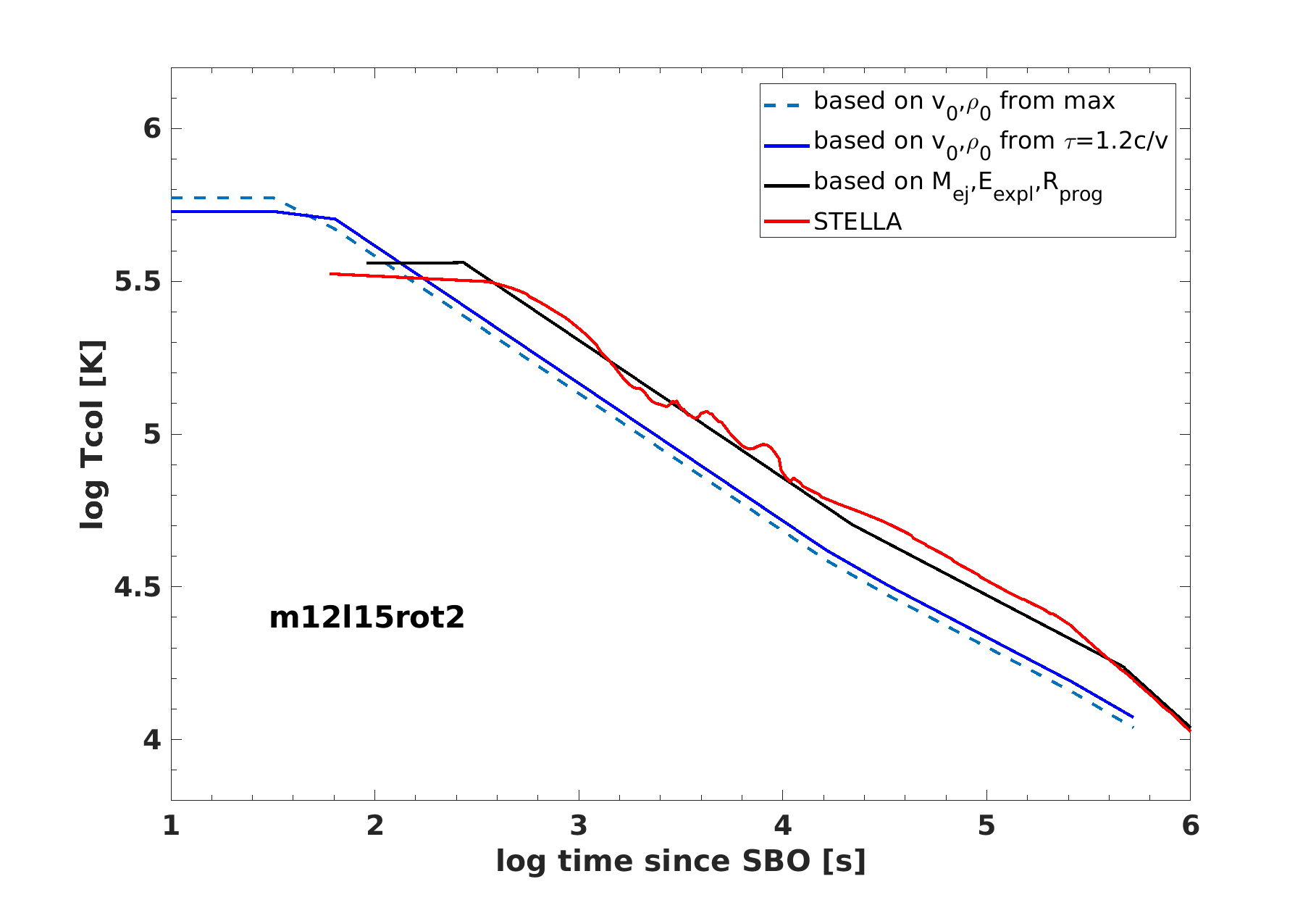}~
\includegraphics[width=0.5\textwidth]{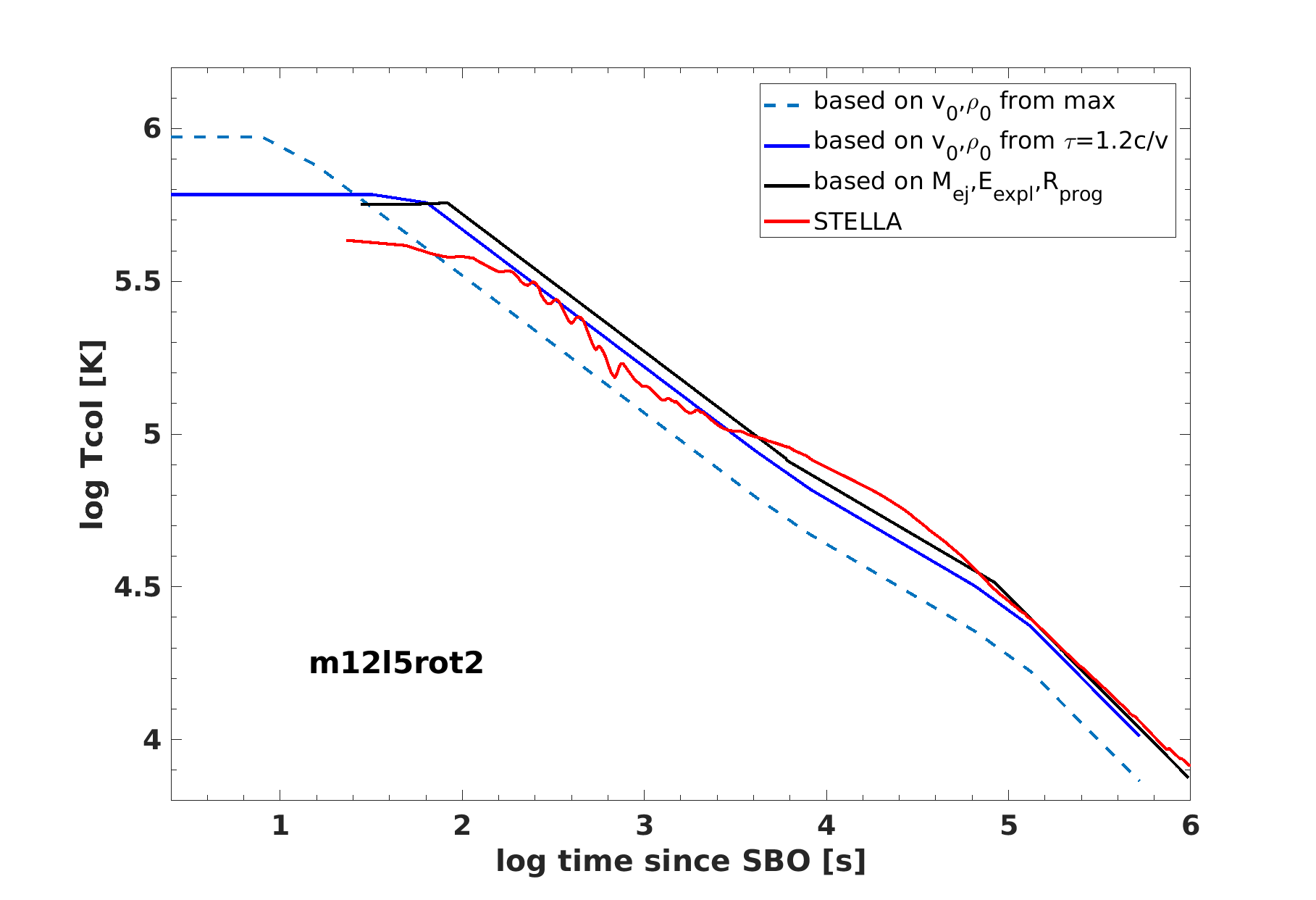}\\
\includegraphics[width=0.5\textwidth]{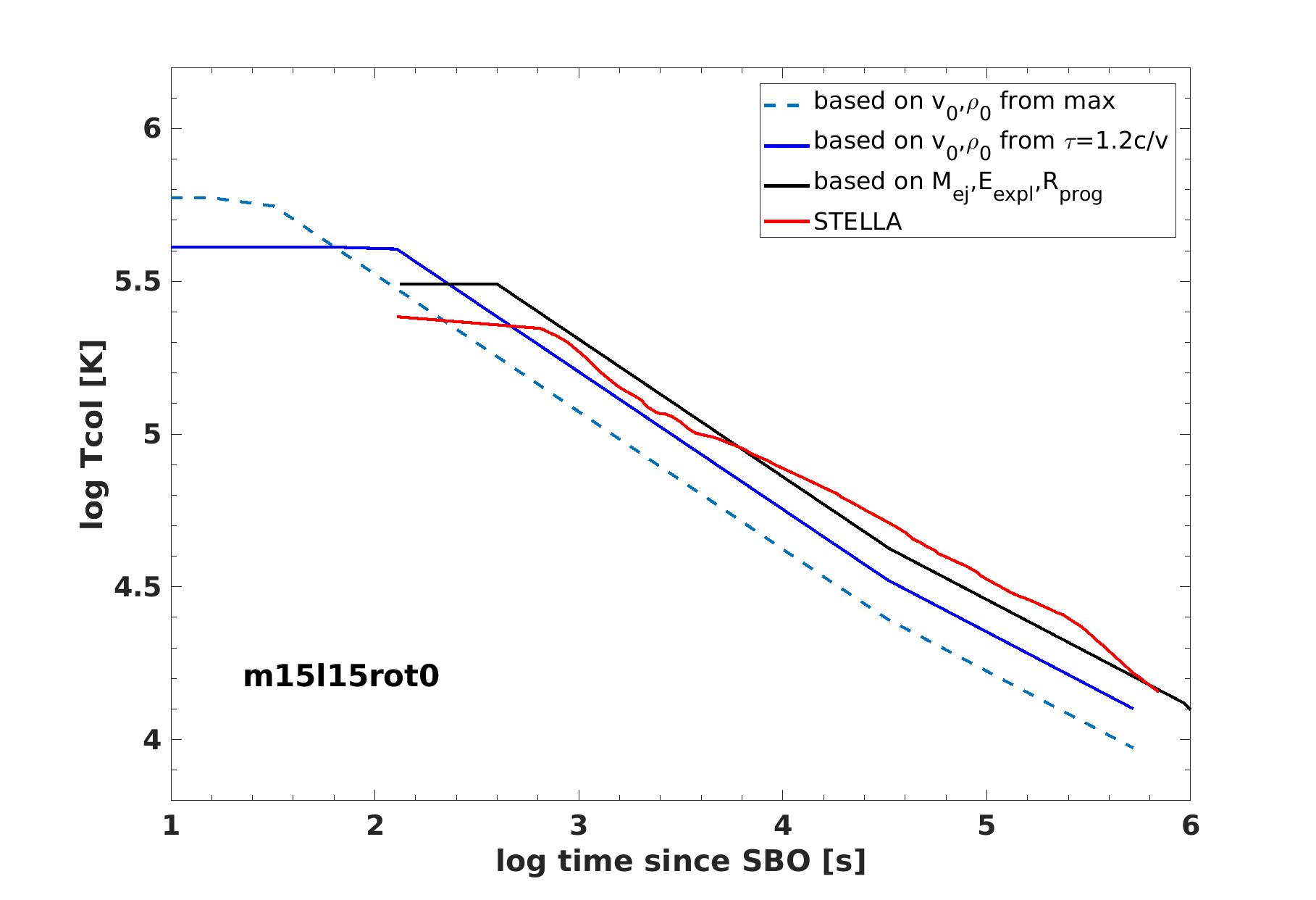}~
\includegraphics[width=0.5\textwidth]{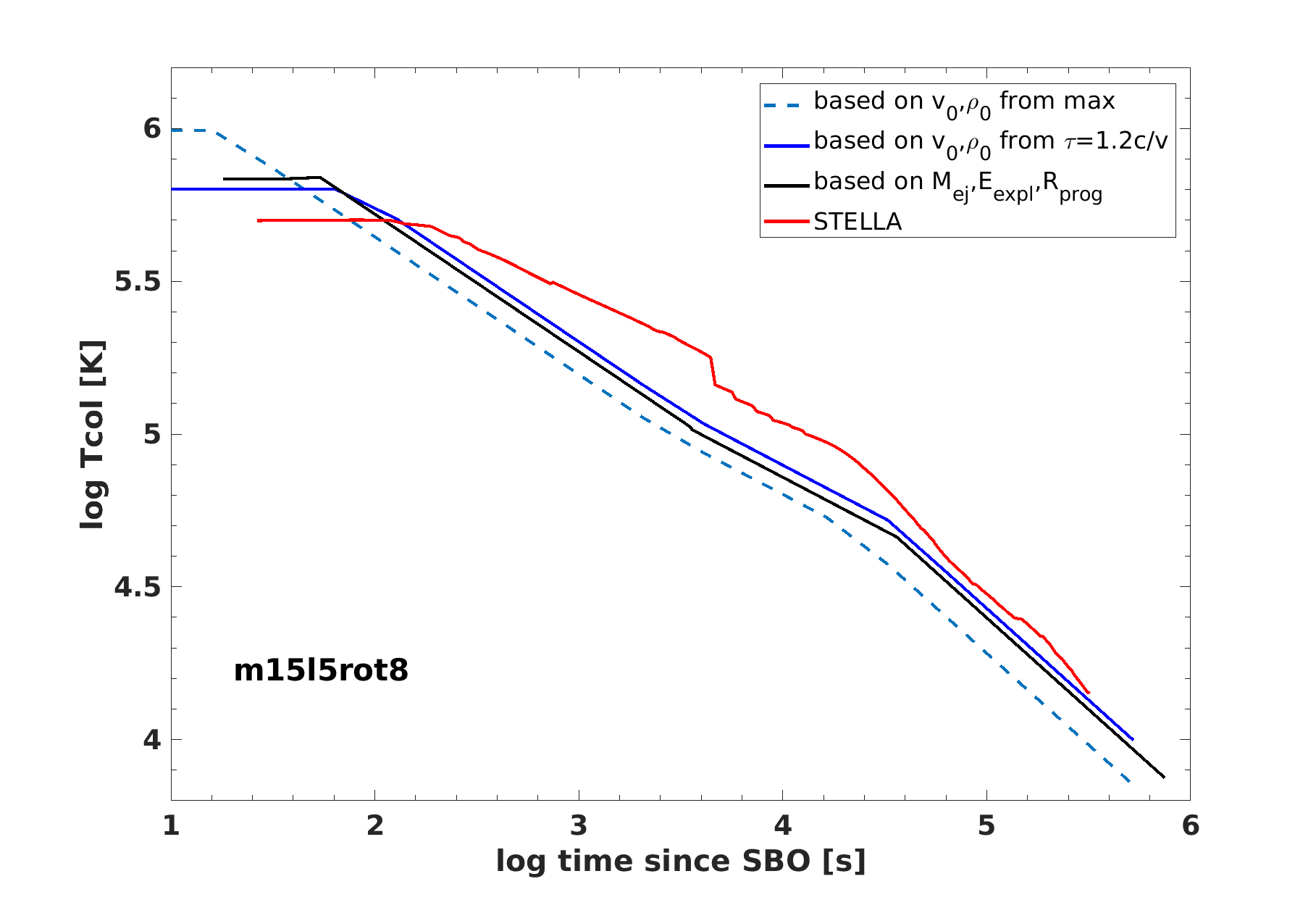}\\
\caption{SBO colour temperature evolution for the red supergiant models in the study.
See details in Figure~\ref{figure:all_Lbol} and in the text.}
\label{figure:all_Tcol}
\end{figure*}

We compare the resulting light curves for the models described in
Section~\ref{sect:method} with those calculated with analytic
formulae from \citet{Shussman2016}. In particular, we used their formulae
(33) and (34), (39) and (41)\footnote{These formulae are implemented
in the web interface \url{http://www.astro.tau.ac.il/~tomersh/}. We list
some formulae from \citet{Shussman2016} which we explore in the current study in Appendix~\ref{appendix:append} for
convenience.}
which define SBO bolometric luminosity and colour temperature from either (1) SBO
parameters, i.e. breakout-shell velocity $v_0$ and density ahead the
shock $\rho_0$ (Eq. 33 and 34), or (2) global supernova parameters, i.e. ejecta mass, radius of the
progenitor and the explosion energy (Eq. 39 and 41, which are implemented on the mentioned
website). The $v_0$ is defined as the maximal shock velocity when the
outermost mass zone reaches 1/3 of the maximal velocity which corresponds to
the moment when outcoming luminosity equals the peak luminosity of the SBO.
As discussed by \citealt{Shussman2016} (see their Section~3.1), 
$v_0$ and $\rho_0$ should be picked from the condition $\tau=1.2 c/v$ instead of
$\tau=c/v_0$, where $\tau$ is Rosseland mean optical depth. We
demonstrate the procedure of the choice in Figure~\ref{figure:VeloTau}.
Here, the red curve corresponds to the profile when the peak luminosity is
reached and when velocity at the surface of the progenitor is 1/3 of the
shock velocity (i.e. maximal velocity).
Figures~\ref{figure:all_Lbol} and \ref{figure:all_Tcol} present the
comparison plots. 
Note that the X-axis in all the plots in the paper is in logarithmic
scale, therefore, the first point of each curve is at time\,=\,0 
relative to the maximum SBO luminosity and could
not be plotted. By default, luminosity and temperature are constant before the
curves begin on the plots, i.e. before 100\,s{}.
We find out that in both cases
peak bolometric luminosity (horizontal part of the dashed and blue curves) 
is overestimated by a factor of at least 0.5~dex if calculated via
Eq.~33 using the values of breakout-shell parameters $v_0$ and $\rho_0$ and
compared to numerically simulated. Nevertheless, the ``analytical'' peak luminosity is very
sensitive to the value of breakout shell velocity (to the exponent of 3) and density
(to the exponent of 1).
Note also, that density just on front of the breakout shell changes dramatically,
and the choice of density value $\rho_0$ strongly affects the resulting peak
luminosity, therefore, it is a source of large uncertainty. On top of that,
calibration done by \cite{Shussman2016} was carried out with the simplified
gray-opacity radiation code, which provides a first-order estimate of the
evolution of radiation field and hydrodynamical evolution tightly coupled with
radiation.
Comparison shows that analytical estimate based on breakout shell parameters 
for the peak luminosity deviates from numerical simulations to 0.5--1~dex.
Apart from that, in Figure~\ref{figure:all_Lbol} and \ref{figure:all_Tcol}, we present analytical
curves calculated from equation (39) from \cite{Shussman2016} using
global progenitor and supernova parameters (black solid curves), i.e. ejecta mass, radius of the
progenitor and the explosion energy, and which are available via website 
\url{http://www.astro.tau.ac.il/~tomersh/}. These analytical curves suit
better the numerical curves deviating from the numerical approach to
0.2~dex, i.e. reaching agreement within a factor of 3.
Maximum breakout luminosity is 
different if compare the present numerical simulations and calibrated analytical values:
$\log L_{bol}^{max}=45.2$\,erg\,s$^{\,-1}$, 45.0\,erg\,s$^{\,-1}$,
45.0\,erg\,s$^{\,-1}$, and 44.9\,erg\,s$^{\,-1}$ in \verb|STELLA| simulations,
while $\log L_{bol}^{max}=45.4$\,erg\,s$^{\,-1}$, 45.4\,erg\,s$^{\,-1}$,
45.3\,erg\,s$^{\,-1}$, and 45.6\,erg\,s$^{\,-1}$ from Eq.~39 (\citealt{Shussman2016})
for the model m12l15rot2, m12l5rot2, m15l15rot0, and m15l5rot8, correspondingly.
Hence, \citet{Shussman2016} calibration overestimates the peak
luminosity what leads to incorrect evaluation of energetics during the
maximum light.

We derive a broken-power
law for the bolometric light curves and colour temperature evolution for our
present simulations. We summarise the values of the average exponents in
Table~\ref{table:slope}. For this, 
each \verb|STELLA| bolometric luminosity curve is divided on 3 linear pieces
in logarithmic scale in accordance with
convenient physical conditions. These are maximum breakout luminosity phase, planar
phase, and so-called spherical phase.
Each piece is approximated by $\log L_{bol} = a + b \log t$, where $t$ is time in
seconds since SBO. 
In Table~\ref{table:slope}, we also present slopes in the luminosity
and colour temperature evolution published in other analytical studes for
comparison.

Colour temperature is analysed similar to bolometric luminosity, i.e. $\log T_{col} = c + d \log t${}.
Maximal temperature $\log T_{col}^{max}$ is in K: 5.5,
5.6, 5.4 and 5.7 in \verb|STELLA|, and 5.6, 5.8, 5.5, and 5.9 from Eq.~41.
We conclude that temporal evolution of SBO luminosity and
colour temperature based on the detailed numerical simulations with
\verb|STELLA| agree well with corresponding evolution presented in \citet{Shussman2016}
while peak values are overestimated in \citet{Shussman2016} in comparison to \verb|STELLA|.

\begin{table*}
\caption{Broken-power law fits for $L_{bol}$ and $T_{col}${} derived from
the present numerical simulations, and the exponents from analytical
formulae from \citet[][]{2010ApJ...725..904N}, \citet{2011ApJ...728...63R},
\citet{2011ApJ...742...36S},
\citet{Shussman2016}, \citet{2017ApJ...838..130S}, and \citet{2019ApJ...879...20F}.
}
\begin{tabular}{|l|c|c|c|c}
\hline
Source &\multicolumn{2}{c}{Lbol}& \multicolumn{2}{c}{Tcol} \\
       & Planar & Spherical & Planar & Spherical \\
       & phase  & phase     & phase  & phase     \\
\hline
\citet{2011ApJ...728...63R} & &0.35/0.16 & & 0.45/0.47\\
\citet{2011ApJ...742...36S} & 4/3 & & & \\
\citet{2017ApJ...838..130S} & & & 0.2 & 0.5 \\
\hline
\citet{2019ApJ...879...20F} & 4/3 &2/3 0.36 & 0.27 & 0.86/0.41/0.6/0.56 \\
\citet{2010ApJ...725..904N} & 4/3& 0.17 & 0.36& 0.56\\
\citet{Shussman2016}        & 4/3& 0.35 & 0.45/0.55 & 0.6\\
\hline
STELLA         & 1.22 & 0.35  & 0.48/0.27 & 0.54 \\
\end{tabular}
\label{table:slope}
\end{table*}

To conclude, correct estimate of the colour temperature provides qualitatively adequate model for
spectral maximum and the integrated flux over the spectral maximum. 
This is very important for the prediction of progenitor parameters from
earlier observations of supernovae.

\subsection[Effect of light travel time]{Effect of light travel time}
\label{subsect:ltt}

A special attention was paid by \cite{Shussman2016} to the effect of light
travel time (see their Section~4.3). This is a geometrical effect which
takes into consideration that photons emitted at an angle $\theta$ to the line of sight
require longer time to travel to the observer. In the extreme case of
$\theta=90^\circ$, the delay is $R/c$, where $R$ is the photon emission radius. 
This leads to smearing of the light curve over a duration of $R/c$. 
STELLA does not account for this effect, consequently, the
numerical results are in reasonable agreement with analytical formulae with no effect
of light travel time included. 
Since the breakout velocity is much lower than the speed of
light, the light travel time affects the light curve significantly only during the initial
light crossing time of the progenitor, namely  $ t \lesssim R_{prog}/c$.  If
one is interested in the light curve during this time then we recommend that
the light travel time would be included.  We found out that there is a
missing factor of $\cos(\theta)$ in the integrand on the right-hand side of
Eq.~22 of \citet[][Section 4.3]{Shussman2016}.  As a result the entire treatment
of the light travel time in \cite{Shussman2016} is mistaken (including
the factor $f$ they finds).  A good approximation of the light travel time
effect is obtained by the following formula:
\begin{equation}
L_{ltt}(t)=\frac{2}{t_{Rc}}\int_{t-t_{Rc}}^t L(t') \left(1-\frac{(t-t')}{t_{Rc}}\right) dt' \, ,
\end{equation}
where $t_{Rc}=R_{prog}/c$. 

For a more accurate treatment see \cite{2012ApJ...747..147K}.
Note that for $t<t_{Rc}$ the integrand on the right-hand side includes also contribution
from $t'<0$ (i.e., before the peak of $L(t)$). This contribution must be taken into account.
An analytic approximation of $L(t<0)$ is given by Eq.~13 of \cite{Shussman2016}.

\subsection[Spectral Energy Distribution]{Spectral Energy Distribution}
\label{subsect:spec}

\begin{figure}
\centering
\includegraphics[width=0.5\textwidth]{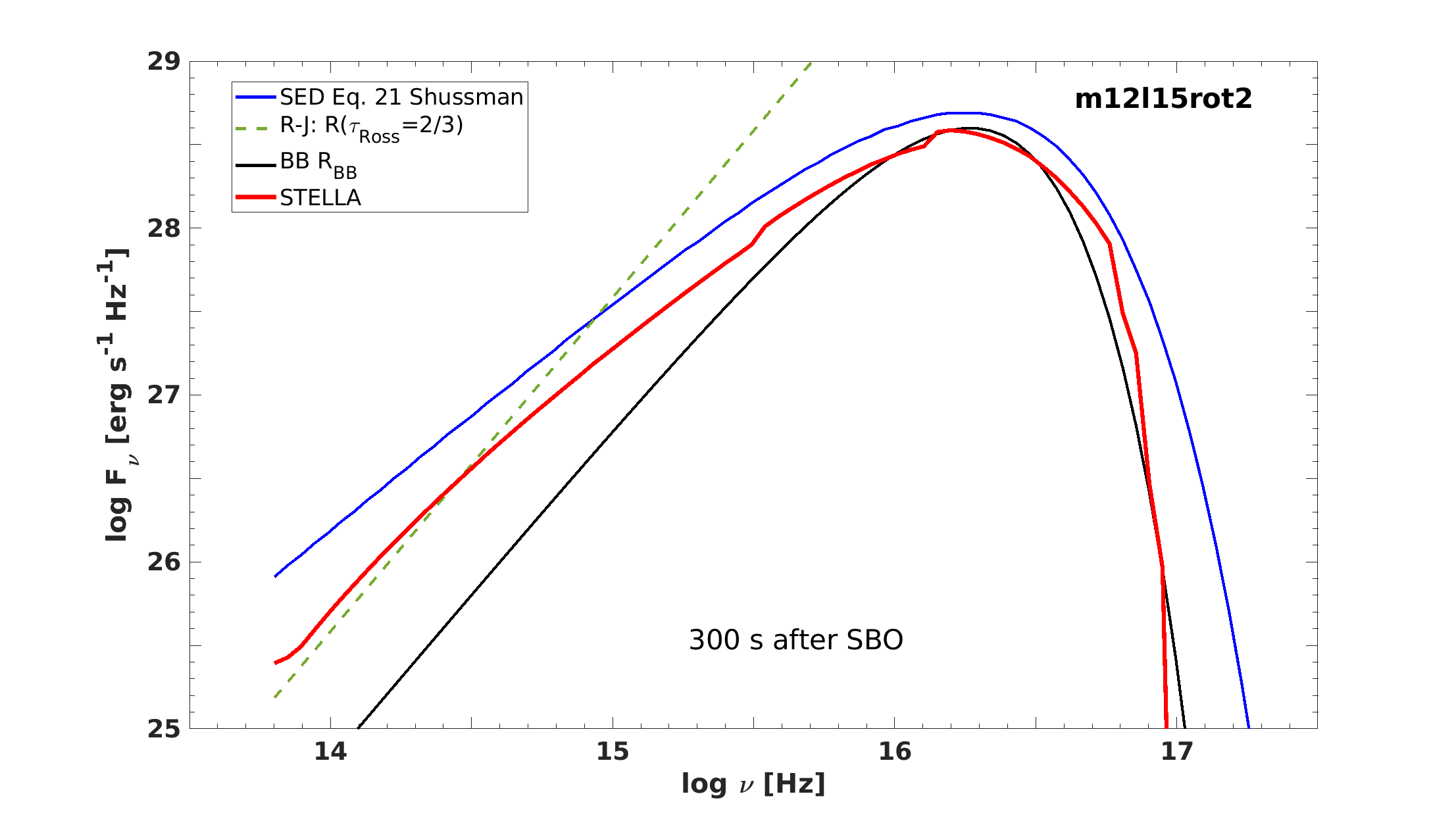}
\caption{Spectral energy distribution for model m12l15rot2 at 300~s after the SBO pulse: computed
with STELLA (red), calculated with analytical formulae from
\citet{Shussman2016} (Eq. 19 and 21, blue),
and the black body SED at colour temperature for the given
moment $T_\mathrm{col}=3.15\times10^{\,5}$~K (black) and Rayleign\,--\,Jeans
formulae (green). See details in the text.}
\label{figure:spec}
\end{figure}

According to Eq.~19 and 21 from \cite{Shussman2016}, the spectral energy
distribution (hereafter, SED) of SBO pulse at given time is:
\begin{equation}
L_\nu=0.9L\cdot{15\over{\pi^{\,4}}}
\left({h\over{3kT_\mathrm{col}(\nu)}}\right)^4 \nu^{\,3} \cdot
\left( e^{h\nu/kT_\mathrm{col}(\nu)} - 1 \right)^{\,-1} \,,
\label{eq21}
\end{equation}
where
\begin{equation}
T_\mathrm{col}(\nu) \approx T_\mathrm{obs} \left({h\nu\over{3kT_\mathrm{obs}}}\right)^{0.2} \, ,
\label{eq19}
\end{equation}
where 
$k=1.38\times10^{\,-16}\,\mathrm{erg\,K}^{\,-1}$ is Boltzmann constant,
$h=6.626\times10^{\,-27}\,\mathrm{erg\,Hz}^{\,-1}$ is Planck constant.
$T_\mathrm{obs}$ and $L=L_\mathrm{obs}$ are generally 
calculated via formulae (33) and (35) or (39) and (41) from
\cite{Shussman2016}.
For demonstration, we apply 
$T_\mathrm{obs} = T_\mathrm{col}^\mathrm{STELLA}$
and $L=L_\mathrm{bol}^\mathrm{STELLA}$.

In Figure~\ref{figure:spec}, the SED for
model m12l15rot2 are taken at 300~s after the SBO pulse. These are: SED from
the \verb|STELLA| numerical simulations (red curve) and analytical spectra (blue curve). The maxima
perfectly coincide, i.e. the colour temperature is estimated correctly in
analytic prescription at this epoch. Additionally we show the black-body SED at emitting radius
$R_{BB}$ (so-called, black-body radius) taken from the \verb|STELLA| results. Pure
black-body underestimates number of low-energy photons because of dilution
in the semi-transparent ejecta \citep{1985Sobolev}. Rayleign\,--\,Jeans
approximation is shown as green dashed curve and taken at radius where
Rosseland optical depth equals 2/3.
In fact, spectrum at a given moment is the sum of
black body spectra at different colour temperatures, i.e. the observer sees
different layers at different frequencies, since optical depth is
frequency dependent. Hence, the resulting spectrum deviates from a pure black body SED
\citep{2011AstL...37..194B}.
The plot shows fairly good agreement between the analytic
prescription Eq.~\ref{eq21} \citep[Eq. 21, ][]{Shussman2016} and
numerical simulations done with \verb|STELLA| around
spectral maximum, as expected. However, we find out that there is a systematic offset of about 0.1~dex between numerical and
analytic curves most likely resulting from the factor 0.9 in Eq.~\ref{eq21} in
\citet[Eq. 21][]{Shussman2016}. 

Among other features, there is a cut-off in numerically computed
spectrum at higher frequencies due to photoionisation (bound-free
transitions) and line-blanketing 
while there is no cut-off in analytically calculated spectrum because the assumptions do not
include influence of photoionisation and effect of line opacity.
 
\begin{figure}
\centering
\includegraphics[width=0.5\textwidth]{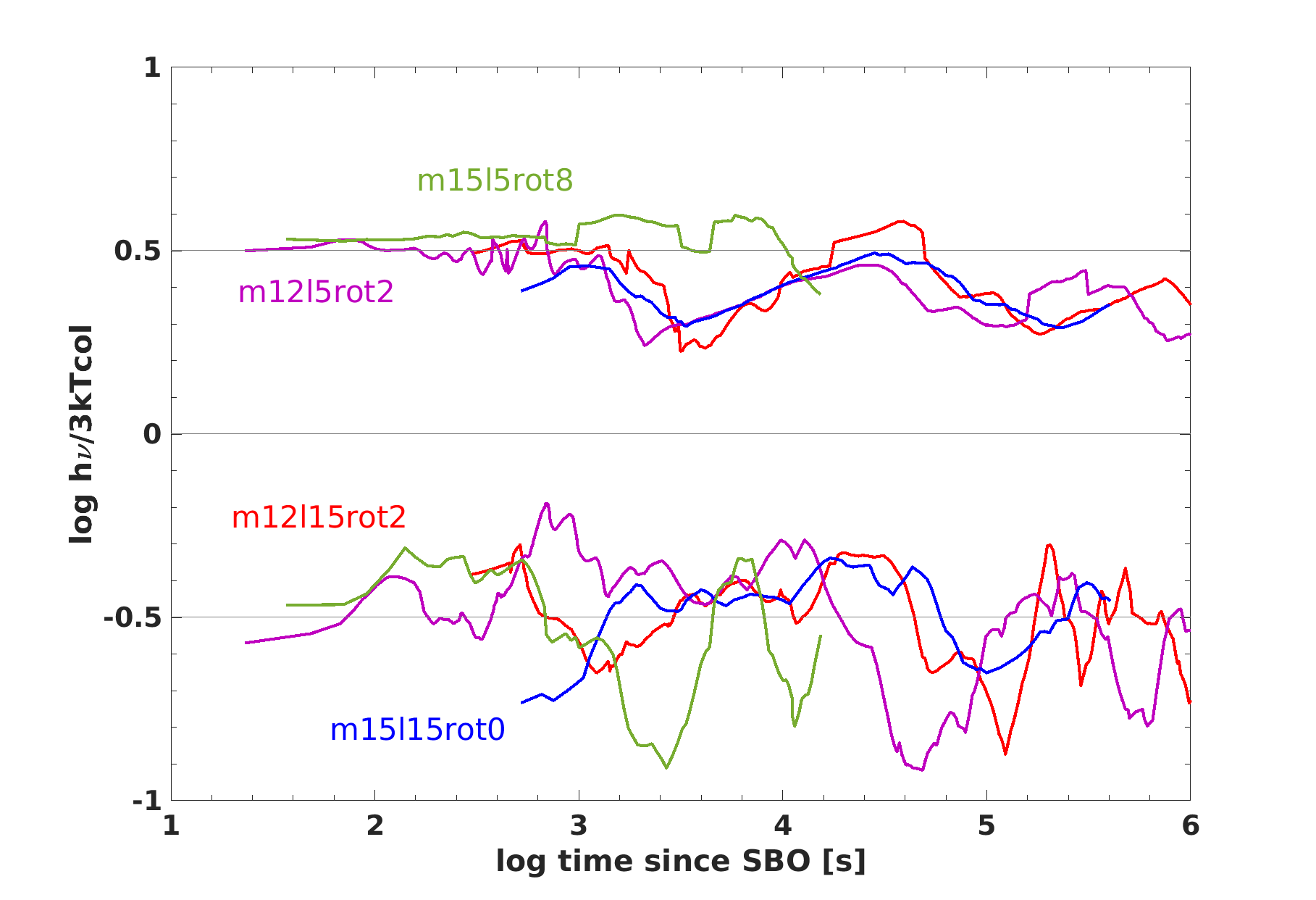}
\caption{Evolution of spectral range (high and low frequency boundaries) in which
analytic formulae from \citet{Shussman2016} are feasible.}
\label{figure:FreqLimits}
\end{figure}

In Figure~\ref{figure:FreqLimits}, we show the upper and lower frequencies
which limit the feasibility range of the given analytical expressions.
We define the limiting frequencies where the difference between analytic SED
deviates from numerical SED less than 1\% (in log-scale), i.e. where
\begin{equation}
{\log F_\nu^\mathrm{an}(\nu) - \log F_\nu^\mathrm{num}(\nu)\over{
\log F_\nu^\mathrm{num}(\nu)}}<0.01\,.
\label{equation:FreqLimit}
\end{equation}
In Figure~\ref{figure:FreqLimits}, the frequency limits are shown for
the quantity $h\nu/k\,T_\mathrm{col}$. The noisy behaviour of the curves is
unavoidable and caused by the numerical nature of the data. Thus, the upper and lower limits for
analytic SED follow colour temperature evolution, so that analytical
formulation of \cite{Shussman2016} is valid in the range of about
$10\times\left({h\nu\over{kT_\mathrm{col}}}\right)$ around spectral maximum.
Major flux is radiated between these limiting frequencies, while there is no
significant contribution at higher frequencies due to line-blanketing and
Compton scattering,
and there is a little contribution at lower frequencies, which can be
estimated from Rayleigh-Jeans formula. In other words, analytical
formulae by \cite{Shussman2016} are valid in the following wavelength range:
\begin{equation}
151.4 \,\textrm{\AA}\, T_5^{\,-1} \, < \, \lambda_\textrm{\AA}\, < \, 1514 \,\textrm{\AA}\,T_5^{\,-1}\,,
\label{equation:limits}
\end{equation}
where $T_5$ is colour temperature in units of $10^{\,5}$\,K, and
$\lambda_\textrm{\AA}$ is wavelength in \AA{}. Hence for colour
temperature of about million degrees at the beginning of the SBO, the validity 
range is between 15\,\AA\, and 151 \,\AA\,, and for temperature
$10^{\,4}$\,K, the range is between 1514\,\AA\, and 15140\,\AA\,.

To conclude, we suggest to compute spectral energy distribution for the SBO
signal as:
\[L_\nu(t)\,= \left\{ \begin{array}{lll}
\mathrm{Rayleign-Jeans}(T_\mathrm{col}) & \mathrm{,\,if} &h\nu/3kT_\mathrm{col} < 0.3 \\
\mathrm{Eq}.~\ref{eq21} & \mathrm{,\,if} &0.3 < h\nu/3kT_\mathrm{col} < 3 \, .
\end{array} \right. \]

\section{Other observational features}
\label{sect:observe}

\subsection{Broad-band light curves}
\label{subsect:broad}

\begin{figure}
\centering
\includegraphics[width=0.5\textwidth]{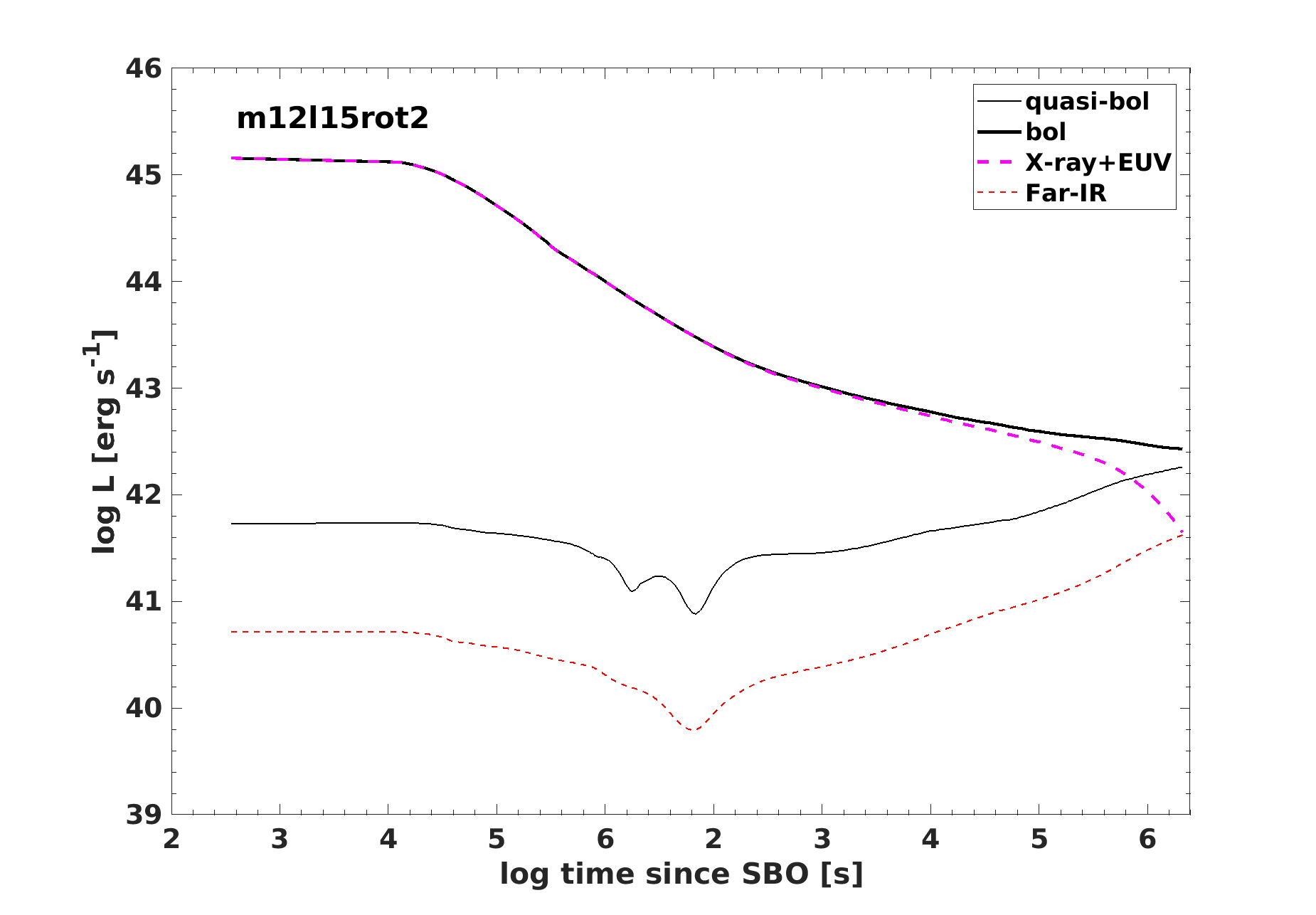}
\caption{Light curves for the model m12l15rot2: bolometric (black thick),
pseudo-bolometric (\emph{UBVRI}, black thin), X-ray$+$Extreme ultraviolet
(3.8~eV\,--\,12.4~keV, magenta
thick dashed), far infra-red ($\lambda>8900$~\AA{}, red thin dashed).}
\label{figure:Xray}
\end{figure}

\begin{figure}
\centering
\includegraphics[width=0.5\textwidth]{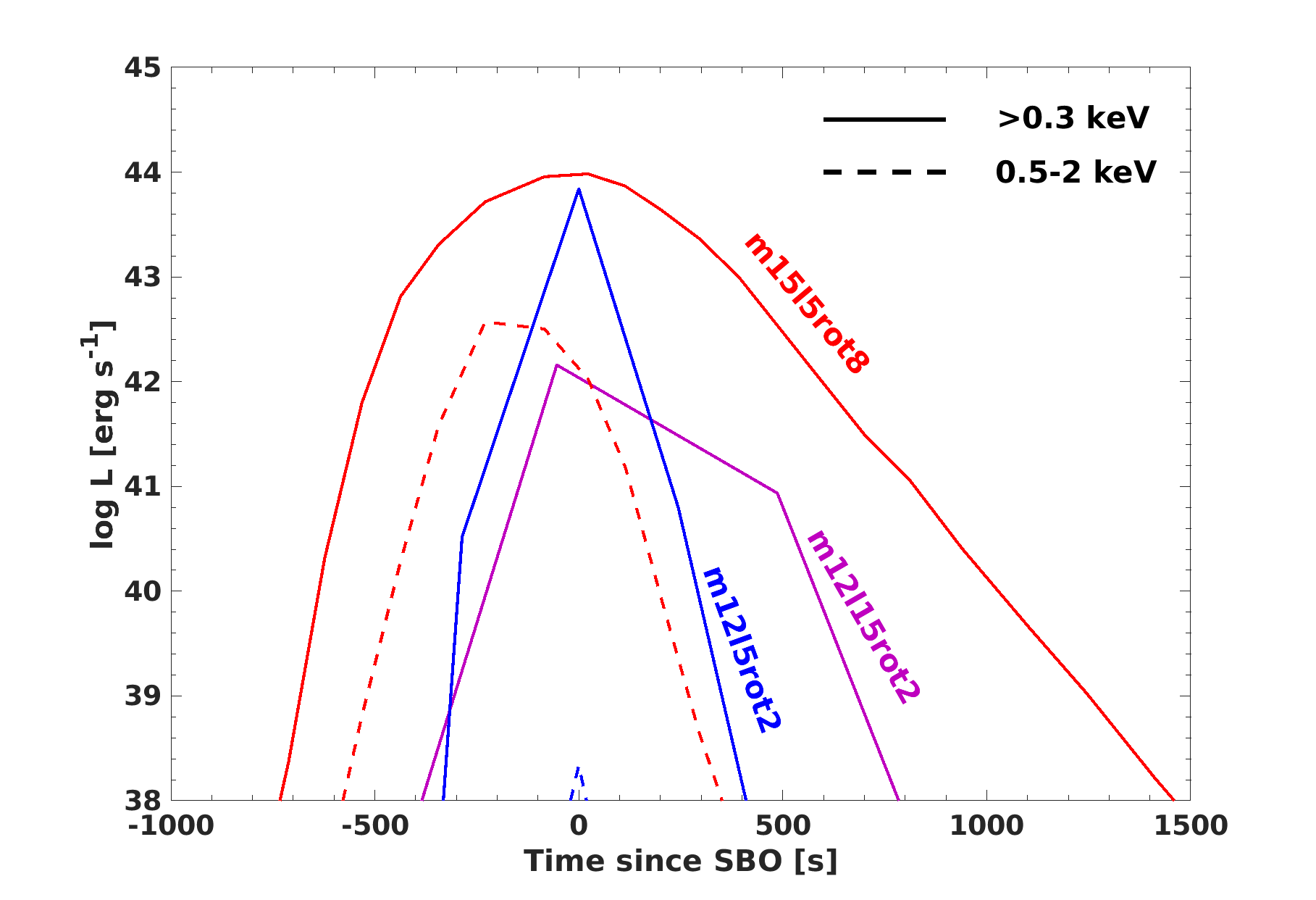}
\caption{Predicted light curves for the models in the study as seen by
eROSITA instrument in the energy range about 0.3~keV (solid) and
0.5\,--2~keV (dashed). The Y-axis limit of
$10^{\,38}$\,erg\,s$^{\,-1}$ is chosen as a typical X-ray luminosity of a normal
star-forming galaxy. Note that the ``0.5-2\,keV''
light curve for the model m12l15rot2 (supposed to be a dashed magenta curve) 
is below the low limit of the plot.}
\label{figure:keV03}
\end{figure}

First of all, we present the overall light curves in different major ranges
of the electromagnetic spectrum.  In Figure~\ref{figure:Xray}, the magenta
curve (X-ray $+$ extreme ultraviolet) represents flux integrated
between 1~\AA{}  and 3250~\AA{} (corresponds to 12.4~keV and 3.8~eV, 
respectively), while the red curve stands for the flux
integrated between 8900~\AA{}  and 50,000~\AA{}. ``Pseudo-bol'' curve (Pseudo-bolometric) stands
for the flux integrated in \emph{UBVRI} spectral range, i.e. between
3250\,\AA {} and 8900\,\AA{}.
Figure~\ref{figure:Xray} demonstrates that
the major flux during SBO comes in the high-energy range, i.e. 
X-ray and ultraviolet flux overwhelmingly contributes to overall bolometric
luminosity. The recently successfully launched mission SPECTRUM-RG
\citep{2012arXiv1209.3114M} observes the sky in the range between 0.3~keV and
30~keV. For comparison, SBOs from normal supernovae, i.e. from exploding
supergiants, have spectral maxima around 0.1~keV at maximum light (soft
X-ray), while compact Wolf-Rayet stars have SBOs in harder X-ray around
1~keV \citep{2010ApJ...725..904N}. Therefore, 
SPECTRUM-RG, particularly, eROSITA instrument 
will detect a moderate SBO signal from normal supernovae with a rate about a
few events during a year (I.\,Khabibulin, R.A.\,Sunyaev, private communication).
Ultraviolet detectors operating around 10~eV might be more optimistic for detecting SBO
from normal supernovae \cite{2014AJ....147...79S}. 
Figure~\ref{figure:keV03} present the predicted light curves for the models in the study
in the eROSITA energy window. We show two options: the solid curves are
those integrated in the energy range about 0.3~keV, while dashed curves are
those integrated in the window 0.5\,--\,2~keV. The later is considered as
more realistic prediction due to a number of reasons, e.g. the exact dependence of the efficient
telescope area on energy. On top on that, the spectrum of the SBO itself varies
dramatically around 0.3\,--\,0.5~keV, and the light curve ``>0.3\,keV'' and
``0.5--2\,keV'' differ significantly. Hence, in some cases (the model
m12l5rot2) the SBO event might be detected if consider flux above 0.3~keV while
it is at the level of the galaxy luminosity in the range 0.5\,--\,2\,keV.
Note that the SED peaks at higher energy for the more compact models since
colour temperature is more sensitive to radius $T_{col}\sim R^{\,-0.46}$
rather than ejecta mass and explosion energy (see the first formula in
Eq.~41). Therefore, the model m15l5rot8 (radius 268~\Rsun{}) and the model m12l15rt2
(345~\Rsun{}) have higher luminosity compared to other models those radii
are 812~\Rsun{} and 1024~\Rsun{}.
We limit the plot with the luminosity $\log L=38$~erg\,s$^{\,-1}$ because this is
a typical X-ray luminosity of a normal star-forming galaxy
\citep{2012MNRAS.426.1870M,2017MNRAS.466.1019S}, and events below $\log L=38$~erg\,s$^{\,-1}$
won't be detected. 

\begin{figure}
\centering
\includegraphics[width=0.5\textwidth]{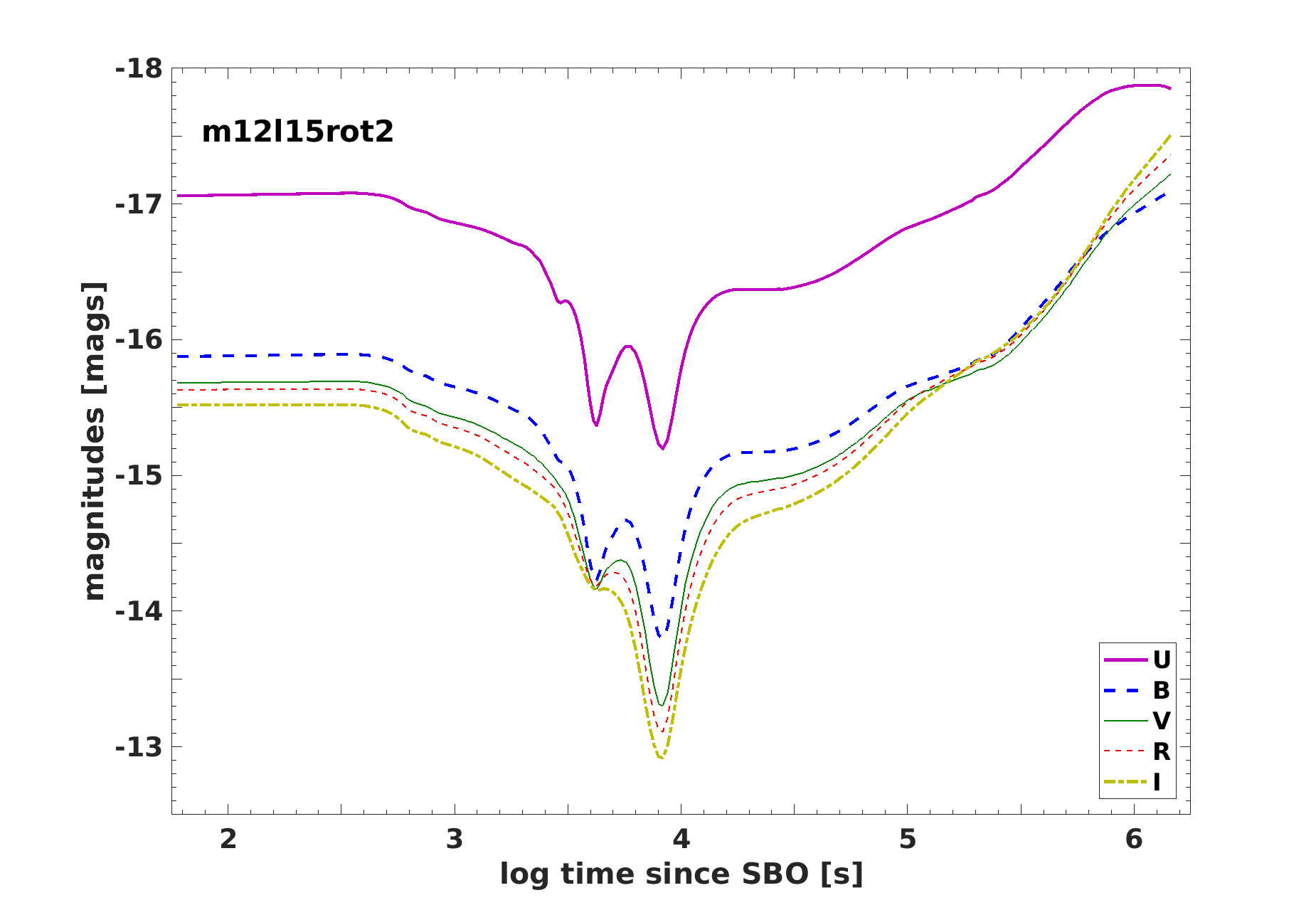}
\caption{Broad-band light curves for the model m12l15rot2.}
\label{figure:ubvri}
\end{figure}

As seen in Figure~\ref{figure:Xray}, luminosity
in visible light and far-infrared range is thousand times lower than in
ultraviolet, and it almost does not contribute to bolometric luminosity,
although it is not negligible. 
Complete photometrical set is unlikely to be observed for the earlier phase,
in particular, SBO pulse, since the survey telescopes usually operate with 1--2 spectral
bands, and the bolometric curve could not be estimated.
Even though the major flux during SBO comes from extreme ultraviolet range,
some noticeable signal might be detected in visible broad bands. 
In Figure~\ref{figure:ubvri}, we show broad-band light curves for the model m12l15rot2.
High-energy flux is dominitating during the first 10~days, therefore,
\emph{U} magnitude is the highest among others. \emph{U} stays at about
$-17^m$ during the first 1000~s while \emph{B},
\emph{V}, \emph{R}, \emph{I} are $-16^m$ for this model. At day~10, i.e. $\sim 10^{\,6}$~s, \emph{U}
peaks at $-18^m$ and then monotonically declines. \emph{B},
\emph{V}, \emph{R}, and \emph{I} rebrighten after 10,000~s, so that the
major flux is radiated in these bands. According to
Figure~\ref{figure:spec} and Figure~\ref{figure:FreqLimits}, 
broad-band magnitudes in visible range might be
easily computed with either the standard Rayleigh-Jeans formula for
$h\nu/3kT_\mathrm{col}< 0.3$, or Equation~\ref{eq21} (i.e. Eq.~(21) from
\citealt{Shussman2016}) for $0.3<h\nu/3kT_\mathrm{col}<3$. The estimate of
magnitudes in optical bands at earlier epoch might serve as an additional constrain to the
progenitor model while analysing the entire set of observational 
data, like light curve, spectra, and photospheric velocity evolution.

\subsection{Photospheric velocity}
\label{subsect:Uphsec}

\begin{figure}
\centering
\includegraphics[width=0.5\textwidth]{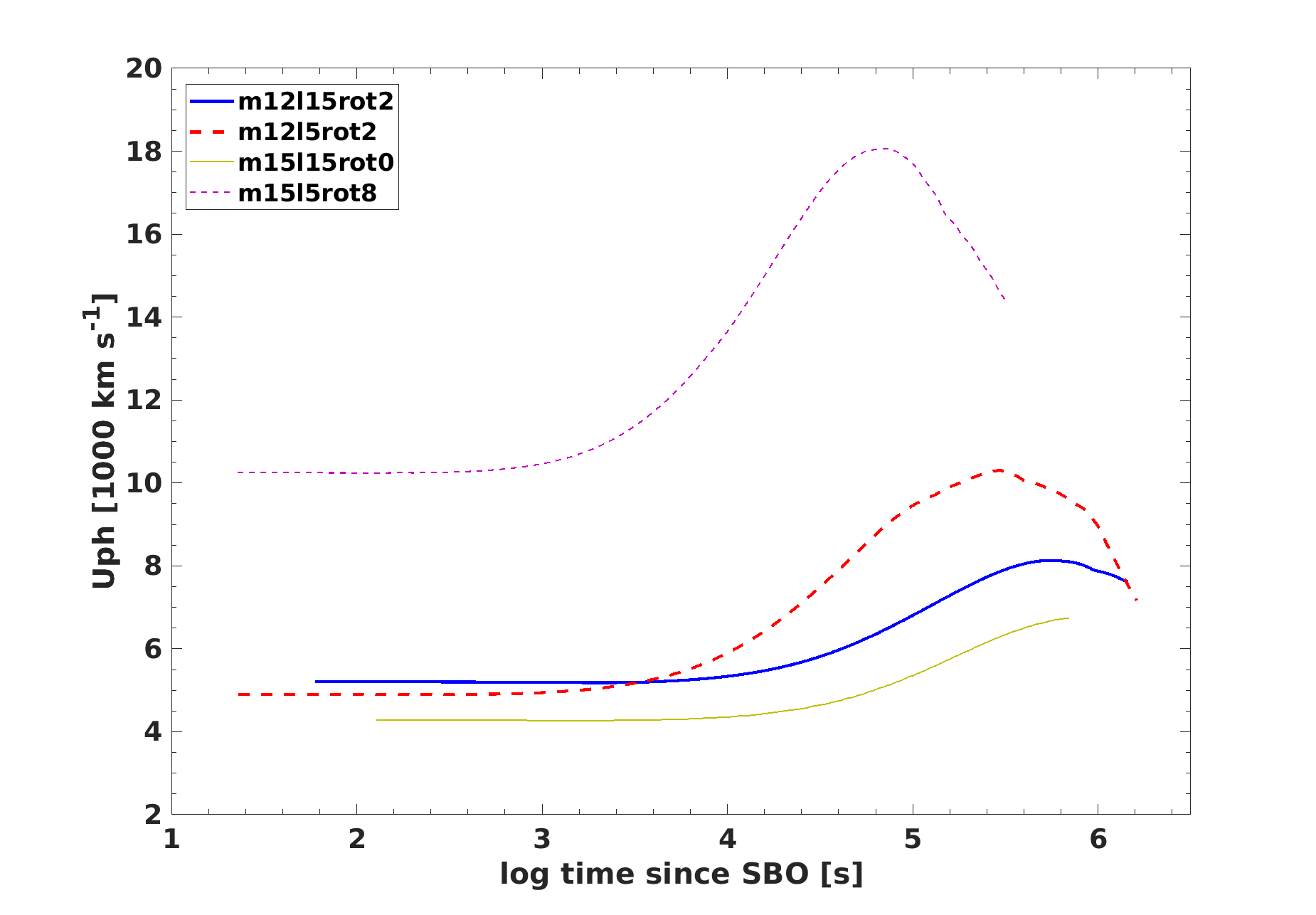}
\caption{Photospheric velocity evolution for the models in the study during
and after SBO.}
\label{figure:Uph}
\end{figure}

Recently, \citet{2019ApJ...879....3G} discussed that the formulae for 
projenitor and explosion parameters derived by
\citet{1993ApJ...414..712P} remain not well defined.
Even being corrected for $^{56}$Ni energy input and different H-to-He ration
in the envelope, Popov's formulation provides
the family of models, i.e. the combination of $M_\mathrm{ej}$,
$R_\mathrm{prog}$ and $E_\mathrm{expl}$, for a given observed SN~IIP light curve. 
Earlier, \citet{2011ApJ...729...61B} also highlighted the question about
degeneracy of hydrodynamical models for a given supernova light curve.
That means that the exact progenitor can not be reliably predicted based on
bolometric observations such as bolometric luminosity and photospheric
velocity. Therefore, additional constrains
from another type of photometric observations are required. \citet{2019ApJ...879....3G} suggest
to use earlier phase observations, in particular, photospheric velocity
observations at day~15, because models in the degenerated family have
essentially different photospheric velocity before day~20 which strongly 
depends on energy of the explosion and ejecta mass (see their Fig. 21).
We show photospheric velocity for the SBO phase in Figure~\ref{figure:Uph}.
Indeed, there is a strong dependence of velocity evolution on mass and
radius of the progenitor. More compact and less massive (in the sense of
ejecta mass) the progenitor,
the higher photospheric velocity at the maximum SBO luminosity. 
Note that explosion energy for all explosions in our study is the same and equal 1~foe. 
According to \citet[][ Eq.~A7]{2010ApJ...725..904N} and \citet[][ Eq.~27a]{Shussman2016}
photospheric velocity at the maximum breakout luminosity is:
\[ v_0\,[\mathrm{km\,s}^{\,-1}]\,\approx \]
\[\left\{ \begin{array}{l}
7,000\,M^{\,-0.43}\,R^{\,-0.26}\,E^{\,0.56}
\approx 7,000\,\left({E\over{M\,R^{\,1/2}}}\right)^{1/2}\,,\\
4,500\,M^{\,-0.44}\,R^{\,-0.49}\,E^{\,0.56}
\approx 4,500\,\left({E\over{M\,R}}\right)^{1/2}\,,
\end{array} \right. \]
correspondingly.

In Figure~\ref{figure:Uph}, it is clearly seen that
measuring velocity at the peak SBO signal might shed light on radius and mass of the progenitor.
The matter is fully ionised at SBO, and spectra, if being recorded,
are featureless. Therefore, photospheric velocity is almost unlikely to be
estimated via spectral observations. However, photospheric velocity could be
calculated from photometric observations via Stefan--Boltzmann law and
a simplified assumption $v=R/t$.

Hence we suggest to use the light curve itself and colour temperature during
the SBO pulse to set additional constrain for
progenitor radius, ejecta mass and energy estimate.

\subsection[Dependence on global supernova parameters]{Dependence on global supernova parameters} 
\label{subsect:depend}

\begin{figure}
\centering
\includegraphics[width=0.5\textwidth]{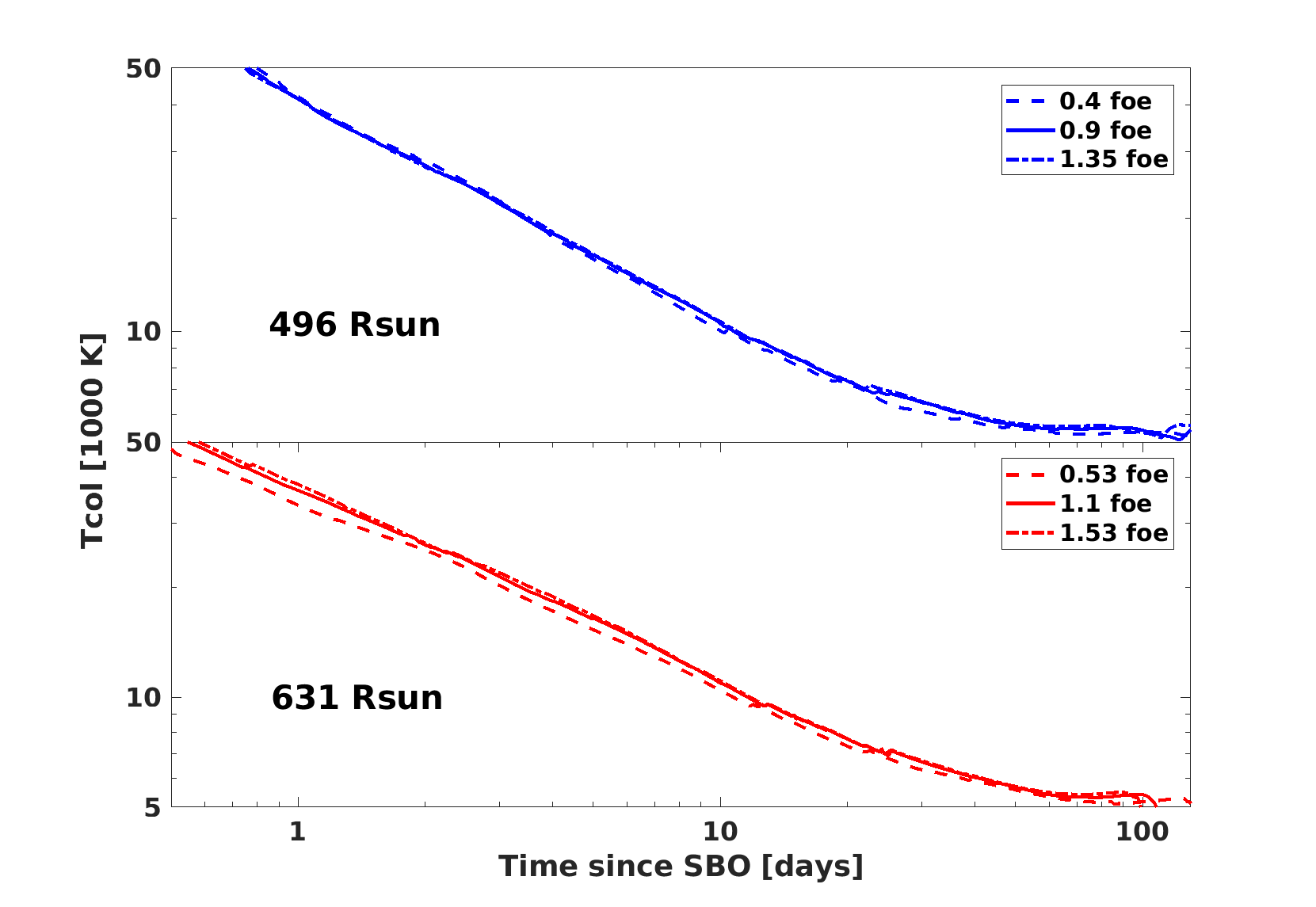}
\caption{Colour temperature for the models m12l3 and m15l3 from
\citep{2019MNRAS.483.1211K} for different explosion energy.}
\label{figure:Edepend}
\end{figure}

We examine the dependence of the colour temperature on explosion energy and
radius based on a wider set of red supergiant models. In Figure~\ref{figure:Edepend},
we show two models from \citet{2019MNRAS.483.1211K} for which explosion
energy scatters in 1~foe, between roughly 0.5~foe and 1.5~foe. This range
for explosion energy is valid for the normal SNe~IIP according to
\citet{2015ApJ...806..225P} and \citet{2017ApJ...841..127M} on the one hand, and also from
the core-collapse explosion simulations by \citet{2016ApJ...818..124E} and
\citet{2016ApJ...821...38S}, on the other hand. To be
precise, the model m12l3 was blown up with 0.4, 0.9 and 1.35~foe, and the
model m15l3 was blown up with 0.53, 1.1 and 1.53~foe. Based on
Figure~\ref{figure:Edepend}, it is clear that colour temperature weakly
depends on explosion energy. We calculate the exponent for the energy dependence as:
\begin{equation}
\alpha={\Delta T_\mathrm{col}/T_\mathrm{col}\over{\Delta E/E}} 
\label{equation:Eexpo}
\end{equation}
using 3 epochs: day~6, 7, 8 and day~4, 5, 6 for the models m12l3 and m15l3,
correspondingly. The averaged value for the exponent is 0.0614, which is 
between 0.027 \citep[Eq. 13, ][]{2011ApJ...728...63R} and 0.11
\citep[Eq. 31, ][]{2010ApJ...725..904N}, and relatively far from --\,0.25
published by \citet[Eq. 41, ][]{Shussman2016}. Nevertheless, all these values
are still pretty close to each other and demonstrate weak dependence of
colour temperature on explosion energy.

\begin{figure}
\centering
\includegraphics[width=0.5\textwidth]{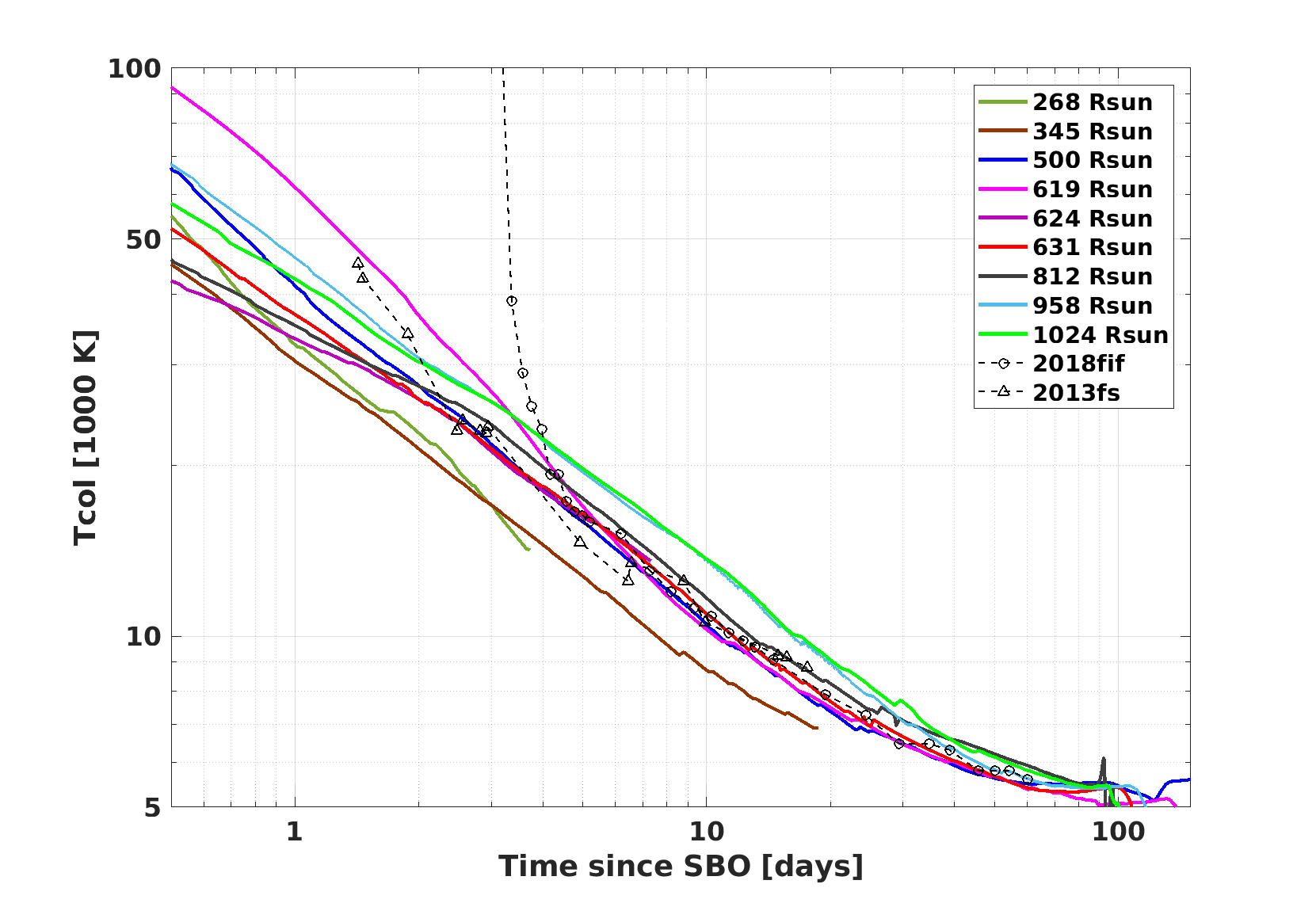}
\caption{Colour temperature for the set of red supergiant models with different radius.
Two observed supernovae are superposed with circles and triangles \citet{2019arXiv190711252S}.}
\label{figure:Rdepend}
\end{figure}

Figure~\ref{figure:Rdepend} shows evolution of colour temperature for a number of red supergiant
models during the so-called spherical phase which precedes the recombination
phase \citep{2014arXiv1404.6313G,2019ApJ...879...20F}. 
These are: 268~\Rsun{} and 345~\Rsun{} - m15l5rot8 and m12l5rot2 from the present study,
500~\Rsun{} and 631~\Rsun{} -- m12l3 and m15l3 from \citet[][]{2019MNRAS.483.1211K}, 
619~\Rsun{} -- L15-pn from \citet{2000ApJS..129..625L}, \citet{2017ApJ...846...37U},
624~\Rsun{} -- 12~\Msun{}-model (unpublished), 812~\Rsun{} -- m12l15rot2 from the present study,
958~\Rsun{} -- hs15.2 from \citet{2016ApJ...818..124E}, \citet{2016ApJ...821...38S},
1024~\Rsun{} -- m15l15rot0 from the present study.  The average ejecta mass is around 10~\Msun{} and the average
explosion energy is around 1~foe. Radius of the progenitors spread over the
large range between 268~\Rsun{} and 1024~\Rsun, and approximately cover the possible
range of radii of red supergiants which explode as SNeIIP 
\citep{2005ApJ...628..973L,2006ApJ...645.1102L,2009MNRAS.395.1409S}.
Note that $T_\mathrm{col}\sim
M_\mathrm{ej}^{\,-\beta}$ where $\beta=0.11$ \citep{2010ApJ...725..904N} or
$\beta=0.13$ \citep{Shussman2016}. Taking into account that
dependence on mass and energy is weak, we calculate the exponent for radius
dependence in a manner similar to the exponent evaluation for energy
dependence. We found that $T_\mathrm{col}\sim
R_\mathrm{prog}^{\,-\gamma}$ where $\gamma=0.386$ which is in very good agreement with
$\gamma=0.38$ \citep{2010ApJ...725..904N} 
and sufficient agreement with $\gamma=0.46$ \citep{Shussman2016}, while
$\gamma=0.25$ in studies by \citet{2011ApJ...728...63R} and \citet{2017ApJ...838..130S}.
We conclude that colour temperature sets a good constrain to
the progenitor radius \citep[see also ][]{2016ApJ...829..109M}. 
However, the accuracy might not be sufficient to
differenciate between progenitors at different initial masses and
metallicities \citep{2017ApJ...848....8R}. Nevertheless, this is related to
a different topic of uncertainty of stellar evolution simulations
\citep[see e.g., ][]{2013A&A...553A..24G,2015MNRAS.447.3115J}.

\section[Conclusions]{Conclusions} 
\label{sect:conclusions}

In the present study, we run four red supergiant models from
\citet{Shussman2016} with the multi-group radiation hydrodynamics code \verb|STELLA|. 
We compare bolometric light curves and colour temperature evolution for
shock breakout and relaxation after the SBO peak to the calibrated
relations published in \citet{Shussman2016}. We found out that analytic
formulae describe reasonably well $L_\mathrm{bol}$ and $T_\mathrm{col}$
evolution during the first day after the shock breakout, however,
overestimate the absolute peak values within a factor of 3. We conclude that the analytic
formulae for SBO pulse provide a fast way to estimate the possible SBO light
curve and colour temperature for a given exploded progenitor, and vice versa, supernova parameters.


We show that there is good agreement between analytically
and numerically computed spectral energy distribution.
We provide a frequency range
(Figure~\ref{figure:FreqLimits}) in which
analytic prescription might be reliably used 
\citet[][ Equations (19) and (21)]{Shussman2016}.


We show that dependence of colour temperature during the spherical phase
on energy is in sufficient agreement with \citet{2011ApJ...728...63R} and 
\citet[][]{2010ApJ...725..904N}, but deviates from \citet{Shussman2016}. The
dependence of colour temperature on progenitor radius is in very good agreement with all three
studes. We suggest to use colour temperature observations prior to recombination phase to
set a constrain on the supernova parameters which is additional to
usual estimates from the late time observations.

We present the prediction for SBO signal from the red supergiants
which might be detected by the recently
launched SPECTRUM-RG/eROSITA X-ray observatory.

\section*{Acknowledgments}
AK and EN are supported by ERC grant No.\,279368
``The Gamma Ray Burst -- Supernova Connection and shock breakout physics''
and partially by the I-Core center of excellence of the CHE-ISF.
AK is also funded by the Alexander von Humboldt Foundation.
SB and PB are sponsored by grant RSF\,18-12-00522. 
The \verb|STELLA| simulations were carried out on the DIRAC Complexity system,
operated by the University of
Leicester IT Services, which forms part of the STFC DiRAC HPC Facility
(\url{www.dirac.ac.uk}). AK thanks Ildar Khabibullin and Adam Rubin for
useful and fruitful discussions.

\addcontentsline{toc}{section}{Acknowledgements}

\bibliographystyle{mnras}
\bibliography{references}

\begin{thebibliography}{}
\makeatletter
\relax
\def\mn@urlcharsother{\let\do\@makeother \do\$\do\&\do\#\do\^\do\_\do\%\do\~}
\def\mn@doi{\begingroup\mn@urlcharsother \@ifnextchar [ {\mn@doi@}
  {\mn@doi@[]}}
\def\mn@doi@[#1]#2{\def\@tempa{#1}\ifx\@tempa\@empty \href
  {http://dx.doi.org/#2} {doi:#2}\else \href {http://dx.doi.org/#2} {#1}\fi
  \endgroup}
\def\mn@eprint#1#2{\mn@eprint@#1:#2::\@nil}
\def\mn@eprint@arXiv#1{\href {http://arxiv.org/abs/#1} {{\tt arXiv:#1}}}
\def\mn@eprint@dblp#1{\href {http://dblp.uni-trier.de/rec/bibtex/#1.xml}
  {dblp:#1}}
\def\mn@eprint@#1:#2:#3:#4\@nil{\def\@tempa {#1}\def\@tempb {#2}\def\@tempc
  {#3}\ifx \@tempc \@empty \let \@tempc \@tempb \let \@tempb \@tempa \fi \ifx
  \@tempb \@empty \def\@tempb {arXiv}\fi \@ifundefined
  {mn@eprint@\@tempb}{\@tempb:\@tempc}{\expandafter \expandafter \csname
  mn@eprint@\@tempb\endcsname \expandafter{\@tempc}}}

\bibitem[\protect\citeauthoryear{{Arcavi} et~al.,}{{Arcavi}
  et~al.}{2017}]{2017ApJ...837L...2A}
{Arcavi} I.,  et~al., 2017, \mn@doi [\apjl] {10.3847/2041-8213/aa5be1}, \href
  {http://adsabs.harvard.edu/abs/2017ApJ...837L...2A} {837, L2}

\bibitem[\protect\citeauthoryear{{Badjin}, {Glazyrin}, {Manukovskiy}  \&
  {Blinnikov}}{{Badjin} et~al.}{2016}]{2016MNRAS.459.2188B}
{Badjin} D.~A.,  {Glazyrin} S.~I.,  {Manukovskiy} K.~V.,   {Blinnikov} S.~I.,
  2016, \mn@doi [\mnras] {10.1093/mnras/stw790}, \href
  {https://ui.adsabs.harvard.edu/abs/2016MNRAS.459.2188B} {459, 2188}

\bibitem[\protect\citeauthoryear{{Bersten}, {Benvenuto}  \& {Hamuy}}{{Bersten}
  et~al.}{2011}]{2011ApJ...729...61B}
{Bersten} M.~C.,  {Benvenuto} O.,   {Hamuy} M.,  2011, \mn@doi [\apj]
  {10.1088/0004-637X/729/1/61}, \href
  {http://adsabs.harvard.edu/abs/2011ApJ...729...61B} {729, 61}

\bibitem[\protect\citeauthoryear{{Blinnikov} \& {Tolstov}}{{Blinnikov} \&
  {Tolstov}}{2011}]{2011AstL...37..194B}
{Blinnikov} S.~I.,  {Tolstov} A.~G.,  2011, \mn@doi [Astronomy Letters]
  {10.1134/S1063773711010051}, \href
  {http://adsabs.harvard.edu/abs/2011AstL...37..194B} {37, 194}

\bibitem[\protect\citeauthoryear{{Blinnikov}, {Eastman}, {Bartunov},
  {Popolitov}  \& {Woosley}}{{Blinnikov} et~al.}{1998}]{1998ApJ...496..454B}
{Blinnikov} S.~I.,  {Eastman} R.,  {Bartunov} O.~S.,  {Popolitov} V.~A.,
  {Woosley} S.~E.,  1998, \mn@doi [\apj] {10.1086/305375}, \href
  {http://adsabs.harvard.edu/abs/1998ApJ...496..454B} {496, 454}

\bibitem[\protect\citeauthoryear{{Blinnikov}, {R{\"o}pke}, {Sorokina},
  {Gieseler}, {Reinecke}, {Travaglio}, {Hillebrandt}  \&
  {Stritzinger}}{{Blinnikov} et~al.}{2006}]{2006A&A...453..229B}
{Blinnikov} S.~I.,  {R{\"o}pke} F.~K.,  {Sorokina} E.~I.,  {Gieseler} M.,
  {Reinecke} M.,  {Travaglio} C.,  {Hillebrandt} W.,   {Stritzinger} M.,  2006,
  \mn@doi [\aap] {10.1051/0004-6361:20054594}, \href
  {http://adsabs.harvard.edu/abs/2006A%26A...453..229B} {453, 229}

\bibitem[\protect\citeauthoryear{{Dessart} \& {Hillier}}{{Dessart} \&
  {Hillier}}{2019}]{2019A&A...625A...9D}
{Dessart} L.,  {Hillier} D.~J.,  2019, \mn@doi [\aap]
  {10.1051/0004-6361/201834732}, \href
  {https://ui.adsabs.harvard.edu/abs/2019A&A...625A...9D} {625, A9}

\bibitem[\protect\citeauthoryear{{Dessart}, {John Hillier}  \&
  {Audit}}{{Dessart} et~al.}{2017}]{2017A&A...605A..83D}
{Dessart} L.,  {John Hillier} D.,   {Audit} E.,  2017, \mn@doi [\aap]
  {10.1051/0004-6361/201730942}, \href
  {https://ui.adsabs.harvard.edu/abs/2017A&A...605A..83D} {605, A83}

\bibitem[\protect\citeauthoryear{{Ensman} \& {Burrows}}{{Ensman} \&
  {Burrows}}{1992}]{1992ApJ...393..742E}
{Ensman} L.,  {Burrows} A.,  1992, \mn@doi [\apj] {10.1086/171542}, \href
  {http://adsabs.harvard.edu/abs/1992ApJ...393..742E} {393, 742}

\bibitem[\protect\citeauthoryear{{Ertl}, {Janka}, {Woosley}, {Sukhbold}  \&
  {Ugliano}}{{Ertl} et~al.}{2016}]{2016ApJ...818..124E}
{Ertl} T.,  {Janka} H.~T.,  {Woosley} S.~E.,  {Sukhbold} T.,   {Ugliano} M.,
  2016, \mn@doi [\apj] {10.3847/0004-637X/818/2/124}, \href
  {https://ui.adsabs.harvard.edu/abs/2016ApJ...818..124E} {818, 124}

\bibitem[\protect\citeauthoryear{{Faran}, {Goldfriend}, {Nakar}  \&
  {Sari}}{{Faran} et~al.}{2019}]{2019ApJ...879...20F}
{Faran} T.,  {Goldfriend} T.,  {Nakar} E.,   {Sari} R.,  2019, \mn@doi [\apj]
  {10.3847/1538-4357/ab218a}, \href
  {https://ui.adsabs.harvard.edu/abs/2019ApJ...879...20F} {879, 20}

\bibitem[\protect\citeauthoryear{{Georgy}, {Ekstr{\"o}m}, {Granada}, {Meynet},
  {Mowlavi}, {Eggenberger}  \& {Maeder}}{{Georgy}
  et~al.}{2013}]{2013A&A...553A..24G}
{Georgy} C.,  {Ekstr{\"o}m} S.,  {Granada} A.,  {Meynet} G.,  {Mowlavi} N.,
  {Eggenberger} P.,   {Maeder} A.,  2013, \mn@doi [\aap]
  {10.1051/0004-6361/201220558}, \href
  {https://ui.adsabs.harvard.edu/abs/2013A&A...553A..24G} {553, A24}

\bibitem[\protect\citeauthoryear{{Gezari} et~al.,}{{Gezari}
  et~al.}{2008}]{2008ApJ...683L.131G}
{Gezari} S.,  et~al., 2008, \mn@doi [\apjl] {10.1086/591647}, \href
  {http://adsabs.harvard.edu/abs/2008ApJ...683L.131G} {683, L131}

\bibitem[\protect\citeauthoryear{{Goldberg}, {Bildsten}  \&
  {Paxton}}{{Goldberg} et~al.}{2019}]{2019ApJ...879....3G}
{Goldberg} J.~A.,  {Bildsten} L.,   {Paxton} B.,  2019, \mn@doi [\apj]
  {10.3847/1538-4357/ab22b6}, \href
  {https://ui.adsabs.harvard.edu/abs/2019ApJ...879....3G} {879, 3}

\bibitem[\protect\citeauthoryear{{Goldfriend}, {Nakar}  \& {Sari}}{{Goldfriend}
  et~al.}{2014}]{2014arXiv1404.6313G}
{Goldfriend} T.,  {Nakar} E.,   {Sari} R.,  2014, arXiv e-prints, \href
  {https://ui.adsabs.harvard.edu/abs/2014arXiv1404.6313G} {p. arXiv:1404.6313}

\bibitem[\protect\citeauthoryear{{Jones}, {Hirschi}, {Pignatari}, {Heger},
  {Georgy}, {Nishimura}, {Fryer}  \& {Herwig}}{{Jones}
  et~al.}{2015}]{2015MNRAS.447.3115J}
{Jones} S.,  {Hirschi} R.,  {Pignatari} M.,  {Heger} A.,  {Georgy} C.,
  {Nishimura} N.,  {Fryer} C.,   {Herwig} F.,  2015, \mn@doi [\mnras]
  {10.1093/mnras/stu2657}, \href
  {http://adsabs.harvard.edu/abs/2015MNRAS.447.3115J} {447, 3115}

\bibitem[\protect\citeauthoryear{{Katz}, {Sapir}  \& {Waxman}}{{Katz}
  et~al.}{2012}]{2012ApJ...747..147K}
{Katz} B.,  {Sapir} N.,   {Waxman} E.,  2012, \mn@doi [\apj]
  {10.1088/0004-637X/747/2/147}, \href
  {https://ui.adsabs.harvard.edu/abs/2012ApJ...747..147K} {747, 147}

\bibitem[\protect\citeauthoryear{{Kozyreva}, {Nakar}  \& {Waldman}}{{Kozyreva}
  et~al.}{2019}]{2019MNRAS.483.1211K}
{Kozyreva} A.,  {Nakar} E.,   {Waldman} R.,  2019, \mn@doi [\mnras]
  {10.1093/mnras/sty3185}, \href
  {http://adsabs.harvard.edu/abs/2019MNRAS.483.1211K} {483, 1211}

\bibitem[\protect\citeauthoryear{{Levesque}, {Massey}, {Olsen}, {Plez},
  {Josselin}, {Maeder}  \& {Meynet}}{{Levesque}
  et~al.}{2005}]{2005ApJ...628..973L}
{Levesque} E.~M.,  {Massey} P.,  {Olsen} K.~A.~G.,  {Plez} B.,  {Josselin} E.,
  {Maeder} A.,   {Meynet} G.,  2005, \mn@doi [\apj] {10.1086/430901}, \href
  {https://ui.adsabs.harvard.edu/abs/2005ApJ...628..973L} {628, 973}

\bibitem[\protect\citeauthoryear{{Levesque}, {Massey}, {Olsen}, {Plez},
  {Meynet}  \& {Maeder}}{{Levesque} et~al.}{2006}]{2006ApJ...645.1102L}
{Levesque} E.~M.,  {Massey} P.,  {Olsen} K.~A.~G.,  {Plez} B.,  {Meynet} G.,
  {Maeder} A.,  2006, \mn@doi [\apj] {10.1086/504417}, \href
  {https://ui.adsabs.harvard.edu/abs/2006ApJ...645.1102L} {645, 1102}

\bibitem[\protect\citeauthoryear{{Limongi}, {Straniero}  \&
  {Chieffi}}{{Limongi} et~al.}{2000}]{2000ApJS..129..625L}
{Limongi} M.,  {Straniero} O.,   {Chieffi} A.,  2000, \mn@doi [\apjs]
  {10.1086/313424}, \href {http://adsabs.harvard.edu/abs/2000ApJS..129..625L}
  {129, 625}

\bibitem[\protect\citeauthoryear{{Litvinova} \& {Nadezhin}}{{Litvinova} \&
  {Nadezhin}}{1985}]{1985SvAL...11..145L}
{Litvinova} I.~Y.,  {Nadezhin} D.~K.,  1985, Soviet Astronomy Letters, \href
  {http://adsabs.harvard.edu/abs/1985SvAL...11..145L} {11, 145}

\bibitem[\protect\citeauthoryear{{Lovegrove}, {Woosley}  \&
  {Zhang}}{{Lovegrove} et~al.}{2017}]{2017ApJ...845..103L}
{Lovegrove} E.,  {Woosley} S.~E.,   {Zhang} W.,  2017, \mn@doi [\apj]
  {10.3847/1538-4357/aa7b7d}, \href
  {http://adsabs.harvard.edu/abs/2017ApJ...845..103L} {845, 103}

\bibitem[\protect\citeauthoryear{{Merloni} et~al.,}{{Merloni}
  et~al.}{2012}]{2012arXiv1209.3114M}
{Merloni} A.,  et~al., 2012, arXiv e-prints, \href
  {https://ui.adsabs.harvard.edu/abs/2012arXiv1209.3114M} {}

\bibitem[\protect\citeauthoryear{{Mineo}, {Gilfanov}  \& {Sunyaev}}{{Mineo}
  et~al.}{2012}]{2012MNRAS.426.1870M}
{Mineo} S.,  {Gilfanov} M.,   {Sunyaev} R.,  2012, \mn@doi [\mnras]
  {10.1111/j.1365-2966.2012.21831.x}, \href
  {https://ui.adsabs.harvard.edu/abs/2012MNRAS.426.1870M} {426, 1870}

\bibitem[\protect\citeauthoryear{{Modjaz} et~al.,}{{Modjaz}
  et~al.}{2009}]{2009ApJ...702..226M}
{Modjaz} M.,  et~al., 2009, \mn@doi [\apj] {10.1088/0004-637X/702/1/226}, \href
  {http://adsabs.harvard.edu/abs/2009ApJ...702..226M} {702, 226}

\bibitem[\protect\citeauthoryear{{Moriya}}{{Moriya}}{2013}]{2013PhDTMoriya}
{Moriya} T.~J.,  2013, PhD thesis, Department of Astronomy, Graduate School of
  Science University of Tokyo, 206 pp.

\bibitem[\protect\citeauthoryear{{Morozova}, {Piro}, {Renzo}  \&
  {Ott}}{{Morozova} et~al.}{2016}]{2016ApJ...829..109M}
{Morozova} V.,  {Piro} A.~L.,  {Renzo} M.,   {Ott} C.~D.,  2016, \mn@doi [\apj]
  {10.3847/0004-637X/829/2/109}, \href
  {https://ui.adsabs.harvard.edu/abs/2016ApJ...829..109M} {829, 109}

\bibitem[\protect\citeauthoryear{{M{\"u}ller}, {Prieto}, {Pejcha}  \&
  {Clocchiatti}}{{M{\"u}ller} et~al.}{2017}]{2017ApJ...841..127M}
{M{\"u}ller} T.,  {Prieto} J.~L.,  {Pejcha} O.,   {Clocchiatti} A.,  2017,
  \mn@doi [\apj] {10.3847/1538-4357/aa72f1}, \href
  {http://adsabs.harvard.edu/abs/2017ApJ...841..127M} {841, 127}

\bibitem[\protect\citeauthoryear{{Nakar} \& {Sari}}{{Nakar} \&
  {Sari}}{2010}]{2010ApJ...725..904N}
{Nakar} E.,  {Sari} R.,  2010, \mn@doi [\apj] {10.1088/0004-637X/725/1/904},
  \href {http://adsabs.harvard.edu/abs/2010ApJ...725..904N} {725, 904}

\bibitem[\protect\citeauthoryear{{Paxton}, {Bildsten}, {Dotter}, {Herwig},
  {Lesaffre}  \& {Timmes}}{{Paxton} et~al.}{2011}]{2011ApJS..192....3P}
{Paxton} B.,  {Bildsten} L.,  {Dotter} A.,  {Herwig} F.,  {Lesaffre} P.,
  {Timmes} F.,  2011, \mn@doi [\apjs] {10.1088/0067-0049/192/1/3}, \href
  {http://adsabs.harvard.edu/abs/2011ApJS..192....3P} {192, 3}

\bibitem[\protect\citeauthoryear{{Paxton} et~al.,}{{Paxton}
  et~al.}{2013}]{2013ApJS..208....4P}
{Paxton} B.,  et~al., 2013, \mn@doi [\apjs] {10.1088/0067-0049/208/1/4}, \href
  {http://adsabs.harvard.edu/abs/2013ApJS..208....4P} {208, 4}

\bibitem[\protect\citeauthoryear{{Paxton} et~al.,}{{Paxton}
  et~al.}{2015}]{2015ApJS..220...15P}
{Paxton} B.,  et~al., 2015, \mn@doi [\apjs] {10.1088/0067-0049/220/1/15}, \href
  {http://adsabs.harvard.edu/abs/2015ApJS..220...15P} {220, 15}

\bibitem[\protect\citeauthoryear{{Paxton} et~al.,}{{Paxton}
  et~al.}{2018}]{2018ApJS..234...34P}
{Paxton} B.,  et~al., 2018, \mn@doi [\apjs] {10.3847/1538-4365/aaa5a8}, \href
  {http://adsabs.harvard.edu/abs/2018ApJS..234...34P} {234, 34}

\bibitem[\protect\citeauthoryear{{Pejcha} \& {Prieto}}{{Pejcha} \&
  {Prieto}}{2015}]{2015ApJ...806..225P}
{Pejcha} O.,  {Prieto} J.~L.,  2015, \mn@doi [\apj]
  {10.1088/0004-637X/806/2/225}, \href
  {http://adsabs.harvard.edu/abs/2015ApJ...806..225P} {806, 225}

\bibitem[\protect\citeauthoryear{{Popov}}{{Popov}}{1993}]{1993ApJ...414..712P}
{Popov} D.~V.,  1993, \mn@doi [\apj] {10.1086/173117}, \href
  {http://adsabs.harvard.edu/abs/1993ApJ...414..712P} {414, 712}

\bibitem[\protect\citeauthoryear{{Rabinak} \& {Waxman}}{{Rabinak} \&
  {Waxman}}{2011}]{2011ApJ...728...63R}
{Rabinak} I.,  {Waxman} E.,  2011, \mn@doi [\apj] {10.1088/0004-637X/728/1/63},
  \href {http://adsabs.harvard.edu/abs/2011ApJ...728...63R} {728, 63}

\bibitem[\protect\citeauthoryear{{Rubin} \& {Gal-Yam}}{{Rubin} \&
  {Gal-Yam}}{2017}]{2017ApJ...848....8R}
{Rubin} A.,  {Gal-Yam} A.,  2017, \mn@doi [\apj] {10.3847/1538-4357/aa8465},
  \href {https://ui.adsabs.harvard.edu/abs/2017ApJ...848....8R} {848, 8}

\bibitem[\protect\citeauthoryear{{Sagiv} et~al.,}{{Sagiv}
  et~al.}{2014}]{2014AJ....147...79S}
{Sagiv} I.,  et~al., 2014, \mn@doi [\aj] {10.1088/0004-6256/147/4/79}, \href
  {https://ui.adsabs.harvard.edu/abs/2014AJ....147...79S} {147, 79}

\bibitem[\protect\citeauthoryear{{Sapir} \& {Waxman}}{{Sapir} \&
  {Waxman}}{2017}]{2017ApJ...838..130S}
{Sapir} N.,  {Waxman} E.,  2017, \mn@doi [\apj] {10.3847/1538-4357/aa64df},
  \href {https://ui.adsabs.harvard.edu/abs/2017ApJ...838..130S} {838, 130}

\bibitem[\protect\citeauthoryear{{Sapir}, {Katz}  \& {Waxman}}{{Sapir}
  et~al.}{2011}]{2011ApJ...742...36S}
{Sapir} N.,  {Katz} B.,   {Waxman} E.,  2011, \mn@doi [\apj]
  {10.1088/0004-637X/742/1/36}, \href
  {https://ui.adsabs.harvard.edu/abs/2011ApJ...742...36S} {742, 36}

\bibitem[\protect\citeauthoryear{{Sazonov} \& {Khabibullin}}{{Sazonov} \&
  {Khabibullin}}{2017}]{2017MNRAS.466.1019S}
{Sazonov} S.,  {Khabibullin} I.,  2017, \mn@doi [\mnras]
  {10.1093/mnras/stw3113}, \href
  {https://ui.adsabs.harvard.edu/abs/2017MNRAS.466.1019S} {466, 1019}

\bibitem[\protect\citeauthoryear{{Schawinski} et~al.,}{{Schawinski}
  et~al.}{2008}]{2008Sci...321..223S}
{Schawinski} K.,  et~al., 2008, \mn@doi [Science] {10.1126/science.1160456},
  \href {http://adsabs.harvard.edu/abs/2008Sci...321..223S} {321, 223}

\bibitem[\protect\citeauthoryear{{Shussman}, {Waldman}  \& {Nakar}}{{Shussman}
  et~al.}{2016}]{Shussman2016}
{Shussman} T.,  {Waldman} R.,   {Nakar} E.,  2016, preprint, \href
  {http://adsabs.harvard.edu/abs/2016arXiv161005323S} {} (\mn@eprint {arXiv}
  {1610.05323})

\bibitem[\protect\citeauthoryear{{Smartt}}{{Smartt}}{2009}]{2009ARA&A..47...63S}
{Smartt} S.~J.,  2009, \mn@doi [\araa] {10.1146/annurev-astro-082708-101737},
  \href {http://adsabs.harvard.edu/abs/2009ARA%26A..47...63S} {47, 63}

\bibitem[\protect\citeauthoryear{{Smartt}, {Eldridge}, {Crockett}  \&
  {Maund}}{{Smartt} et~al.}{2009}]{2009MNRAS.395.1409S}
{Smartt} S.~J.,  {Eldridge} J.~J.,  {Crockett} R.~M.,   {Maund} J.~R.,  2009,
  \mn@doi [\mnras] {10.1111/j.1365-2966.2009.14506.x}, \href
  {https://ui.adsabs.harvard.edu/abs/2009MNRAS.395.1409S} {395, 1409}

\bibitem[\protect\citeauthoryear{{Sobolev}}{{Sobolev}}{1985}]{1985Sobolev}
{Sobolev} V.~V.,  1985, Moscow Izdatel Nauka, \href
  {https://ui.adsabs.harvard.edu/abs/1985MoIzN....Q....S} {p. 504~pp.}

\bibitem[\protect\citeauthoryear{{Soderberg} et~al.,}{{Soderberg}
  et~al.}{2008}]{2008Natur.453..469S}
{Soderberg} A.~M.,  et~al., 2008, \mn@doi [\nat] {10.1038/nature06997}, \href
  {https://ui.adsabs.harvard.edu/abs/2008Natur.453..469S} {453, 469}

\bibitem[\protect\citeauthoryear{{Soumagnac} et~al.,}{{Soumagnac}
  et~al.}{2019}]{2019arXiv190711252S}
{Soumagnac} M.~T.,  et~al., 2019, arXiv e-prints, \href
  {https://ui.adsabs.harvard.edu/abs/2019arXiv190711252S} {p. arXiv:1907.11252}

\bibitem[\protect\citeauthoryear{{Sukhbold}, {Ertl}, {Woosley}, {Brown}  \&
  {Janka}}{{Sukhbold} et~al.}{2016}]{2016ApJ...821...38S}
{Sukhbold} T.,  {Ertl} T.,  {Woosley} S.~E.,  {Brown} J.~M.,   {Janka} H.-T.,
  2016, \mn@doi [\apj] {10.3847/0004-637X/821/1/38}, \href
  {http://adsabs.harvard.edu/abs/2016ApJ...821...38S} {821, 38}

\bibitem[\protect\citeauthoryear{{Tolstov}, {Blinnikov}  \&
  {Nadyozhin}}{{Tolstov} et~al.}{2013}]{2013MNRAS.429.3181T}
{Tolstov} A.~G.,  {Blinnikov} S.~I.,   {Nadyozhin} D.~K.,  2013, \mn@doi
  [\mnras] {10.1093/mnras/sts577}, \href
  {http://adsabs.harvard.edu/abs/2013MNRAS.429.3181T} {429, 3181}

\bibitem[\protect\citeauthoryear{{Tominaga}, {Morokuma}, {Blinnikov},
  {Baklanov}, {Sorokina}  \& {Nomoto}}{{Tominaga}
  et~al.}{2011}]{2011ApJS..193...20T}
{Tominaga} N.,  {Morokuma} T.,  {Blinnikov} S.~I.,  {Baklanov} P.,  {Sorokina}
  E.~I.,   {Nomoto} K.,  2011, \mn@doi [\apjs] {10.1088/0067-0049/193/1/20},
  \href {http://adsabs.harvard.edu/abs/2011ApJS..193...20T} {193, 20}

\bibitem[\protect\citeauthoryear{{Utrobin}, {Wongwathanarat}, {Janka}  \&
  {M{\"u}ller}}{{Utrobin} et~al.}{2017}]{2017ApJ...846...37U}
{Utrobin} V.~P.,  {Wongwathanarat} A.,  {Janka} H.-T.,   {M{\"u}ller} E.,
  2017, \mn@doi [\apj] {10.3847/1538-4357/aa8594}, \href
  {http://adsabs.harvard.edu/abs/2017ApJ...846...37U} {846, 37}

\makeatother
\end{thebibliography}
\appendix
\section[Analytical formulae]{Analytical formulae from Shussman\,et\,al.\,2016}
\label{appendix:append}

\subsection{Bolometric light curve and observed temperature as functions of the breakout properties
(Section~5.1, Shussman\,et\,al., 2016)}

In Equation 33, 35, and 42 below we use velocity of the SBO shell prior to SBO $v\equiv
v_{0,5}$ in units 5,000~km\,s$^{\,-1}${}, density $\rho\equiv \rho_{0,-9}$ in units
10$^{\,-9}$\,g\,cm$^{\,-3}${}, and radius of the progenitor prior to SBO $R\equiv R_{500}$ in
units 500~\Rsun{}.\\

Equation~\textbf{33}:
\[ L_\mathrm{obs}(t)\,[\mathrm{erg\,s}^{\,-1}]\simeq \]
\[\left\{ \begin{array}{cllllc}
1.6\times10^{\,45}& v^{\,3}   & \rho          & R^{\,2}    &                  & t<t_0 \\
3.2\times10^{\,43}& v^{\,0.33}& \rho^{\,-0.33}& R^{\,-1.69}& t_{hr}^{\,-4/3}  & t_0\ll t\ll t_s \\
3.3\times10^{\,42}& v^{\,1.31}& \rho^{\,-0.33}& R^{\,0.71} & t_{day}^{\,-0.35}& t_s\ll t<t_{rec} 
\end{array} \right. \] 

Equation~\textbf{35}:
\[ T_\mathrm{obs}(t)\,[\mathrm{K}]\simeq \]
\[\left\{ \begin{array}{cllllc}
4.2\times10^5& v^{\,0.76} & \rho^{\,0.24} &           &                 & t<t_0 \\
1.1\times10^5& v^{\,-0.13}& \rho^{\,-0.21}& R^{\,-0.1}& t_{hr}^{-0.45}  & t_0\le t<t_s \\
3.3\times10^4& v^{\,-0.03}& \rho^{\,-0.2} & R^{\,-0.2}& t_{day}^{-0.35} & t_s\le t<t_c \\
4.1\times10^4& v^{\,-0.55}& \rho^{\,0.18} & R^{\,0.06}& t_{day}^{-0.6}  & t_c\le t<t_{rec} 
\end{array} \right. \] 
Here, $t_0$, $t_s$, $t_c$, and $t_{rec}$ are so-called transition times.
$t_0$ marks the end of maximum SBO luminosity phase and the beginning of 
planar phase, $t_s$ is the end of planar phase and the beginning of 
spherical phase, and $t_{rec}$ is the end of spherical phase and the beginning of
recombination phase. $t_c$ is so-called
``colour'' transition time, when thermalization point reaches the breakout
shell and as a consequence colour temperature undergoes a break in its
temporal evolution during spherical phase \citep[Section\,3.2, ][]{Shussman2016}.
We list the expressions for the transition times below in 
Equations \textbf{36a}, \textbf{36b}, \textbf{36c}, and \textbf{36d}:
\[\begin{array}{llll}
t_0=190\,{\rm s}      &v^{\,-2}   & \rho^{\,-1}   & R^{\,-0.23} \\
t_s=3.2\,{\rm hr}     &v^{\,-1}   &               & R           \\
t_c=2.5\,{\rm days}   &v^{\,-2.07}& \rho^{\,0.08} & R^{\,1.06}  \\
t_{rec}=17\,{\rm days}&v^{\,-0.92}& \rho^{\,-0.31}& R^{\,0.1}
\end{array}\]

\subsection{Bolometric light curve and observed temperature as functions of global SN properties
(Section~5.2)}

In the equations below we use radius of the progenitor prior to SBO $R\equiv R_{500}$ in
units 500~\Rsun{}, ejecta mass $M\equiv M_{15}$ in units 15~\Msun{}, energy of the
explosion $E\equiv E_{51}$ in units 10$^{\,51}$~erg.\\

Equation~\textbf{39}:
\[ L_\mathrm{obs}(t)\,[\mathrm{erg\,s}^{\,-1}]\simeq \]
\[\left\{ \begin{array}{rlllll}
1.8\times10^{\,45}& M^{\,-0.65} & R^{\,-0.11}& E^{\,1.37} &                & t\ll t_0 \\
2.7\times10^{\,43}& M^{\,-0.34} & R^{\,1.74} & E^{\,0.29} & t_{\rm hr}^{\,-4/3}  & t_0\ll t\ll t_s \\
1.6\times10^{\,42}& M^{\,-0.78} & R^{\,0.28} & E^{\,0.84} & t_{\rm day}^{\,-0.35}& t_s\ll t<t_{rec} 
\end{array} \right. \] 

Equation~\textbf{41}:
\[ T_\mathrm{obs}(t)\simeq \]
\[\left\{ \begin{array}{rlllll}
4.3\times10^5& M^{\,-0.17} & R^{\,-0.52}& E^{\,0.35}  &                 & t<t_0 \\
1\times10^5  & M^{\,-0.07} & R^{\,0.1}  & E^{\,-0.01} & t_{\rm hr}^{\,-0.45}  & t_0\le t<t_s \\
3\times10^4  & M^{\,-0.11} & R^{\,-0.04}& E^{\,0.04}  & t_{\rm day}^{\,-0.35} & t_s\le t<t_c \\
4.1\times10^4& M^{\,0.13}  & R^{\,0.46} & E^{\,-0.25} & t_{\rm day}^{\,-0.6}  & t_c\le t<t_{rec} 
\end{array} \right. \] 

Equations~\textbf{42a}, \textbf{42b}, \textbf{42c}, \textbf{42d}:
\[\begin{array}{llll}
t_0=155\,{\rm s}        &M^{\,0.23} &R^{\,1.39} &E^{\,-0.81}\\
t_s=3.6\,{\rm hr}       &M^{\,0.44} &R^{\,1.49} &E^{\,-0.56}\\
t_c=3.2\,{\rm days}     &M^{\,0.97} &R^{\,2.02} &E^{\,-1.19}\\
t_{rec}=16.6\,{\rm days}&M^{\,0.22} &R^{\,0.76} &E^{\,-0.43}
\end{array}\]

\bsp    
\label{lastpage}
\end{document}